\documentclass[prd,twocolumn,showpacs,reprint,preprintnumbers,nofootinbib,amsmath,amssymb]{revtex4-2}

\RequirePackage[colorlinks=true
,urlcolor=blue
,anchorcolor=blue
,citecolor=blue
,filecolor=blue
,linkcolor=blue
,menucolor=blue
,linktocpage=true
,pdfproducer=medialab
,pdfa=true,
]{hyperref}

\usepackage{amsmath,amssymb,bm,graphicx,xcolor,hyperref}
\usepackage[capitalize]{cleveref}

\newcommand{\GeV}{\,\mathrm{GeV}}

\newcommand{\keV}{\,\mathrm{keV}}

\usepackage{breqn} 
\makeatletter 
\let\cref@old@eq@setnumber\eq@setnumber 
\def\eq@setnumber{
\cref@old@eq@setnumber
\cref@constructprefix{equation}{\cref@result}
\protected@xdef\cref@currentlabel{
[equation][\arabic{equation}][\cref@result]\p@equation\theequation}} 
\makeatother 

\usepackage{cases}

\crefname{section}{Sec.}{Secs.}
\crefname{figure}{Fig.}{Figs.}
\crefname{equation}{Eq.}{Eqs.}
\crefname{appendix}{Appendix}{Appendices}

\newcommand{\be}{\begin{equation}\begin{aligned}}
\newcommand{\ee}{\end{aligned}\end{equation}}

\newcommand{\beq}{\begin{equation}}
\newcommand{\eeq}{\end{equation}}
\newcommand{\beqa}{\begin{eqnarray}}
\newcommand{\eeqa}{\end{eqnarray}}

\renewcommand{\eqref}[1]{Eq.~(\ref{#1})}

\newcommand{\eg}{e.g.}

\RequirePackage[normalem]{ulem}

\DeclareUnicodeCharacter{2212}{\textendash}

\usepackage{bbold}
\usepackage[OT1]{fontenc}

\begin{document}

\title{Freeze-in of anapole and charge-radius inelastic dark matter: \\X-ray probes of a decoupled vector portal}

\author{Krzysztof Jod\l{}owski}
\email{kjodlowski@njnu.edu.cn}
\email{krzysztofjjodlowski@gmail.com}

\author{Chih-Ting Lu}
\email{ctlu@njnu.edu.cn}

\affiliation{Department of Physics and Institute of Theoretical Physics \\Nanjing Normal University\char`,{} Nanjing\char`,{} 210023\char`,{} China}
\affiliation{Nanjing Key Laboratory of Particle Physics and Astrophysics\char`,{} Nanjing\char`,{} 210023\char`,{} China}

\date{\today}

\begin{abstract}
    We study freeze-in production and experimental probes of inelastic dark matter (iDM) coupled to the Standard Model through dimension-6 transition anapole and charge-radius operators.
    These interactions arise upon integrating out the heavy dark photon of a broken $U(1)_D$ vector portal.
    We focus on mass splittings below the dielectron threshold, where the excited-state decay is cosmologically long-lived and decays predominantly through $\chi_1\to\chi_0+3\gamma$, producing a distinctive keV--MeV photon spectrum.
    We calculate the decay rate and photon spectrum analytically in the Euler--Heisenberg limit and evaluate finite-electron-mass corrections using the full one-loop amplitude.
    We then combine X-ray limits with freeze-in relic density targets and complementary direct detection (DD) and accelerator constraints.
    The surviving $\chi_1$ population also leads to exothermic downscattering signatures at DD experiments, including the possibility of explaining the LZ signal, while the three-photon decay is constrained by X-ray data.
    At $\Delta=900\,\mathrm{keV}$, INTEGRAL/SPI probes the Boltzmann-suppressed high-mass branches of the freeze-in targets, whereas the NuSTAR limits at $\Delta=100\,\mathrm{keV}$ are substantially weaker because of the steep splitting dependence of the decay width.
    Our results demonstrate the potential of X-ray observations as a probe of freeze-in through a decoupled vector portal and of the reheating temperature governing the iDM production.
\end{abstract}

\maketitle

\section{Introduction}
\label{sec:intro}
Identifying how dark matter (DM) is produced and what its physical properties are remains one of the central open problems in particle physics and cosmology~\cite{Bergstrom:2000pn,Bertone:2004pz,Planck:2018vyg}. The comprehensive experimental program to search for DM, including direct detection (DD), indirect detection (ID), and collider searches, is largely motivated by the thermal WIMP paradigm~\cite{Arcadi:2017kky,Roszkowski:2017nbc}. In particular, freeze-out relic density targets have been a key driver of the DM search program, and continue to be a major focus of current and future experiments~\cite{Billard:2021uyg,Arcadi:2024ukq}. However, the absence of a conclusive signal~\cite{Arcadi:2024ukq} motivates the exploration of alternative DM scenarios, especially those leading to distinctive experimental signatures in current or near-future experiments.

Among the many well-motivated DM production mechanisms, freeze-in taking place at low reheating temperature is particularly interesting in this respect~\cite{Cosme:2023xpa,Boddy:2024vgt,Bernal:2024yhu}, as it can connect intensity frontier (IF) searches to the early thermal history.
Establishing the experimental targets for freeze-in DM is important for guiding the developing DM and long-lived particle (LLP) search program, a prime target of which are renormalizable portals that connect the Standard Model (SM) and the DM by a new mediating particle.

A simple realization is a vector portal in which a massive dark photon (DP), associated with a $U(1)_D$ gauge symmetry, mixes kinetically with Standard Model (SM) hypercharge and couples to dark fermions.
When the DS is non-minimal, \eg, when there are two Weyl fermions with opposite charges under $U(1)_D$, and mass terms that both preserve and violate this symmetry, the DM consists of two mass eigenstates, the stable $\chi_0$ and the decaying $\chi_1$ -- a scenario called inelastic DM (iDM).
In the literature, the regime when the DM states and the mediators (DP and, in case the $U(1)_D$ gauge symmetry is spontaneously broken by its vacuum expectation value, the dark Higgs (DH)) are both in the MeV--GeV range is comprehensively studied using a diverse set of signatures~\cite{Alexander:2016aln,Darme:2017glc,Fabbrichesi:2020wbt,Ferber:2023iso,Krnjaic:2025zjl}.
However, the scenario when the mediators are much heavier than iDM species is less explored - a gap our work partly fills.

More concretely, we show that when DP is heavy and decouples, the pseudo-Dirac iDM regime leads to dominant interactions of DM with a single photon given by the charge radius (CR) operator. 
We also discuss pseudo-Majorana iDM regime, when the phenomenology is determined by transition anapole moment (AM) operator instead.
For the first time, we determine their relic abundance resulting from freeze-in, with special focus on the low reheating temperature scenarios.
We explore the prospects of experimental probing of them at DD, ID searches utilizing X-rays, and at the IF.

In the pseudo-Dirac regime, a small splitting is technically natural because it is controlled by the DH Yukawa couplings whose vanishing restores the symmetry.
For theoretical and phenomenological reasons, we concentrate on the regime in which the mass splitting lies below the dielectron threshold.
In such case, the excited DM component typically has a lifetime exceeding the age of the Universe, as its dominant decay channel is the semi-visible three-photon decay, $\chi_1\to\chi_0 +3\gamma$, which is both loop and phase-space suppressed.
At present, this process results in a continuous photon spectrum with energies in the $\sim $keV--MeV range, accessible to X-ray searches like INTEGRAL/SPI~\cite{Bouchet:2011fn,Cirelli:2020bpc} and NuSTAR~\cite{NuSTAR_tech_desc}.
We analytically determine the resulting photon spectrum at the leading order (LO), called in this context the Euler--Heisenberg (EH) limit~\cite{Heisenberg:1936nmg}, while the general case is treated numerically by implementing the 1-loop virtual photon decay within Package-X~\cite{Patel:2015tea,McDermott:2017qcg}.
For AM iDM, we recast the bounds determined in recent papers analyzing signal of iDM with light DP, INTEGRAL/SPI~\cite{Krnjaic:2025zjl} and NuSTAR~\cite{Jeesun:2026ryo}; for CR iDM, we find that the spectrum is given by the same function as in~\cite{Krnjaic:2025zjl,Jeesun:2026ryo}, thus, we apply these limits directly.
For splittings close to the dielectron threshold, these observations probe the high-mass, Boltzmann-suppressed branches of the freeze-in relic-density targets.

Finally, we discuss complementary probes of our model at DD and IF searches.
Since in our model the excited DM component is cosmologically long-lived, there is a large number density of it at present, allowing the kinematically unsuppressed exothermic (downscattering) process. 
In particular, we examine a possible CR interpretation of the recent LZ recoil candidate.
Moreover, decays of vector mesons, dominant at beam dump experiments, and the Drell--Yan process at the LHC, allow an efficient production of both stable and excited DM states, leading to scattering signatures involving both endothermic and exothermic reactions.
We focus on the SND@LHC detector~\cite{SHiP:2020sos,Ahdida:2750060}, covering a region complementary to the forward detectors such as FASER$\nu$~\cite{FASER:2019dxq,FASER:2020gpr} and FLArE~\cite{Batell:2021blf}.

The paper is organized as follows. 
In \cref{sec:model}, we introduce and discuss our model.
In \cref{sec:decay}, we discuss the main signature of our scenario, the semi-visible decay of the excited DM component into the stable DM component and three photons.
In \cref{sec:freezein}, we describe the details of the freeze-in relic density calculation, which is used to establish the target for X-ray observations, DD and IF searches.
In \cref{sec:Bounds}, we describe the current and future bounds due to iDM decays or scatterings.
In \cref{sec:results}, we present our main result: the contours corresponding to the correct DM relic density and the parameter-space coverage from existing and future data.
Finally, we summarize in \cref{sec:conclusions}, while the DH matching and transition Rayleigh phenomenology are given in Appendix~\ref{app:Ray}.

\section{The model}
\label{sec:model}

\subsection{iDM with couplings to single or two photons}
\label{sec:iDM_EMFF}

iDM contains two nearly degenerate states, denoted $\chi_0$ and $\chi_1$, with mass splitting $\Delta=m_{\chi_1}-m_{\chi_0}\ll m_{\chi_0}$.
Small splittings arise in several UV-motivated dark sectors, including supersymmetric models~\cite{TuckerSmith:2001hy} and the pseudo-Dirac limit of the broken $U(1)_D$ model discussed below.
As a result, both the cosmological evolution and the observable signatures differ substantially from those obtained for the elastic coupling.

At energies below the electroweak (EW) scale, the leading off-diagonal interaction Lagrangian between the stable and the excited DM states with a single photon can be written as follows:
\begin{align}
    \mathcal L_{\rm int}^\gamma
    =\,&\frac{i}{2\Lambda_E}\,\bar\chi_1[\gamma^\mu,\gamma^\nu]\gamma^5\chi_0 \,F_{\mu\nu} +\frac{1}{2\Lambda_M}\,\bar\chi_1[\gamma^\mu,\gamma^\nu]\chi_0 \,F_{\mu\nu} 
    \nonumber\\
    &+a_\chi \,\bar\chi_1\gamma^\mu\gamma^5\chi_0\,\partial^\nu F_{\mu\nu}
    +i\,b_\chi\,\bar\chi_1\gamma^\mu\chi_0\,\partial^\nu F_{\mu\nu}
\label{eq:L_int}
\,,
\end{align}
where the first two terms are dimension-5 transition electric and magnetic dipoles (EDM/MDMs), while the last two correspond to the AM and CR operators of dimension-6; all coefficients are taken to be real.
On the other hand, couplings of iDM to two-photon start only from dimension-7 transition Rayleigh operators,
\begin{align}
    \mathcal L_{\rm int}^{\gamma\gamma} =\,&  
    C_S\,\bar\chi_1\chi_0 \,F_{\mu\nu}F^{\mu\nu} +C_P\,\bar\chi_1 i\gamma^5\chi_0 \, F_{\mu\nu}\widetilde F^{\mu\nu} \nonumber\\
    &+\widetilde C_S\, \bar\chi_1\chi_0 \, F_{\mu\nu}\widetilde F^{\mu\nu} +\widetilde C_P\, \bar\chi_1i\gamma^5\chi_0 \, F_{\mu\nu}F^{\mu\nu}
\label{eq:L_int_2}
\,,
\end{align}
where for Majorana mass eigenstates with the same CP parity, the operators with coefficients $C_S$ and $C_P$ are CP-even, while those with $\widetilde C_S$ and $\widetilde C_P$ are CP-odd.

The operators given in \cref{eq:L_int,eq:L_int_2} are higher-dimensional,  requiring an UV-complete origin.
For example, loops of heavy charged particles can generate electromagnetic interactions of a decoupled dark sector~\cite{Weiner:2012gm}.
Depending on the symmetry properties of the renormalizable interactions, one may generate any such operator, or possibly several at the same time.
When transition dipoles are generated with comparable suppression scales, they typically dominate the phenomenology, as discussed in Refs.~\cite{Dienes:2023uve,Fortin:2011hv,Sigurdson:2004zp}.

Instead, in this work, we study higher-dimensional operators: AM and CR dimension-6 transition interactions, and two of dimension-7 Rayleigh transition operators, $C_S$ and $\widetilde C_P$.
The dimension-6 transition operators arise at tree level upon integrating out the heavy dark photon, while DH exchange induces the dimension-7 Rayleigh operators with coefficients $C_S$ and $\widetilde C_P$ through the Higgs portal and the loop-induced scalar coupling to photons.
We investigate benchmarks for which this contribution is subdominant.
As complementary results, we derive the Rayleigh two-photon decay widths, spectra, and X-ray constraints in Appendix~\ref{app:Ray}.

In both cases, the excited DM component is long-lived, since it decays to two or three photons, and the resulting photon spectrum is peaked near $E_\gamma\simeq\Delta/2$ or $E_\gamma\simeq\Delta/3$, for transition Rayleigh and AM/CR operators, respectively.
This setup connects the early-Universe processes responsible for achieving the correct DM relic density with present-day X-ray signatures and searches at terrestrial experiments, resulting in  an economical, testable framework.

\subsection{UV completion for AM and CR iDM}
\label{sec:UV}
We consider a broken, anomaly-free $U(1)_D$ gauge symmetry and two left-handed Weyl fermions $\eta$ and $\xi$ with charges $+1$ and $-1$, respectively, and a complex scalar $\Phi$ with charge $-2$.
The Lagrangian is
\be
    \mathcal{L} &= \mathcal{L}_{\mathrm{SM}} + \mathcal{L}_{A'} + \mathcal{L}_{\Phi} + \mathcal{L}_{f} 
\,, 
\ee
where
\be
\label{eq:L_setup}
    \mathcal{L}_{A'} &= -\frac{1}{4}F'_{\mu\nu}{F'}^{\mu\nu} -\frac{\epsilon}{2\cos\theta_W}F'_{\mu\nu}B^{\mu\nu} 
    \,,
    \\
    \mathcal{L}_{\Phi} &= (D^\mu \Phi)^\dagger (D_\mu \Phi)+ \mu_{\Phi}^2|\Phi|^2-\frac{\lambda_\Phi}{2}|\Phi|^4-\lambda_{H\Phi}|\Phi|^2|H|^2
    \,,
    \\
    \mathcal{L}_{f}  &\supset g_D A'_\mu J_D^\mu -m_D\,\eta\,\xi -\frac{1}{2} \left( y_\eta\Phi\,\eta\eta +y_\xi\Phi^\dagger\xi\xi \right) +\mathrm{h.c.}
    \,,
\ee
where the fermion terms were written in a CP-conserving basis and the parameters are real.
After $U(1)_D$ breaking, we write $\Phi=(v_D+\phi)/\sqrt{2}$ in unitary gauge, giving the DP mass $m_{A'}=2g_D v_D$.
The charge-two condensate leaves an unbroken discrete gauge symmetry $\mathbb Z_2$, under which $\eta$ and $\xi$ are odd, ensuring the stability of $\chi_0$.
After the EW symmetry breaking, $B_{\mu\nu}=\cos \theta_W F_{\mu\nu}-\sin \theta_W Z_{\mu\nu}$, where $\theta_W$ is the weak mixing angle.
The symmetry-breaking contributions to the fermion masses are
\begin{equation}
    f_\eta=y_\eta\langle\Phi\rangle
\,,
    \qquad f_\xi=y_\xi\langle\Phi\rangle
\,.
\end{equation}
We also define
\begin{equation}
    \bar f=\frac{f_\eta+f_\xi}{2}\,, 
    \qquad \delta f=f_\eta-f_\xi
\,.
\end{equation}
Then, the fermion mass matrix and its eigenvalues are
\begin{equation}
\mathcal{M} =
\begin{pmatrix}
    f_\eta & m_D\\
    m_D & f_\xi
    \end{pmatrix}
    \,,
    \quad
    \lambda_\pm=\bar f\pm r
    \,,
    \quad
    r=\sqrt{m_D^2+\frac{(\delta f)^2}{4}}
\,.
\end{equation}
The mixing angle can be chosen such that
\begin{equation}
    \sin(2\theta)=\frac{m_D}{r}
    \,,
    \qquad
    \cos(2\theta)=\frac{\delta f}{2r}
\,.
\end{equation}
After rephasing negative-mass eigenstates and ordering the physical states as $m_{\chi_1}>m_{\chi_0}$, the fermionic dark current is
\begin{equation}
    J_D^\mu = \sin(2\theta) J_{\mathrm{inel.}}^\mu +\frac{\cos(2\theta)}{2} \left( \bar\chi_1\gamma^\mu\gamma^5\chi_1 -\bar\chi_0\gamma^\mu\gamma^5\chi_0 \right)
\,.
\end{equation}
The transition current is determined by the sign of the determinant of the mass matrix:
\begin{equation}
    J_{\mathrm{inel.}}^\mu =
    \begin{cases}
    \displaystyle \pm\bar\chi_1\gamma^\mu\gamma^5\chi_0
    \,, & \operatorname{sgn}(\det\mathcal{M})=+1
    \,,
    \\[6pt]
    \displaystyle
    \pm i\bar\chi_1\gamma^\mu\chi_0
    \,, & \operatorname{sgn}(\det\mathcal{M})=-1
    \,.
\end{cases}
\end{equation}
Thus, $\det\mathcal{M}>0$ gives a transition axial current, whereas $\det\mathcal{M}<0$ gives a transition vector current.  
The boundary $\det\mathcal{M}=0$ contains a massless eigenstate and lies outside the regime considered here.

We work in the regime in which the DP and DH can be integrated out.
We first discuss consequences of integrating out $A'_\mu$, which results in generating AM or CR operators, while in the Appendix~\ref{app:Ray}, we describe the consequences of integrating out $\Phi$, which instead result in two-photon operators $C_S$ and $\widetilde C_P$.

After decoupling heavy $A'_\mu$, to leading order in $\epsilon$ and $q^2/m_{A'}^2$, the effective Lagrangian is
\begin{align}
    \mathcal{L}_{\mathrm{eff}}^{\gamma}
    \supset{}&
    \frac{\epsilon g_D}{m_{A'}^2}\sin(2\theta) J_{\mathrm{inel.}}^\mu\partial^\nu F_{\mu\nu}\\ &+ \frac{\epsilon g_D}{2m_{A'}^2}\cos(2\theta) \left( \bar\chi_1\gamma^\mu\gamma^5\chi_1 -\bar\chi_0\gamma^\mu\gamma^5\chi_0 \right) \partial^\nu F_{\mu\nu}
\,.  
\nonumber
\end{align}
The tree-level matching results in AM and CR operators, given by the second line in \cref{eq:L_int}, which have the following coefficients:
\begin{equation}
    (a_\chi,b_\chi) = \frac{\epsilon g_D}{m_{A'}^2}\sin(2\theta)
    \begin{cases}
    (\pm1,0), & \operatorname{sgn}(\det\mathcal{M})=+1\,,
    \\[4pt]
    (0,\pm1), & \operatorname{sgn}(\det\mathcal{M})=-1
\,,
    \end{cases}
\end{equation}
so small values of $a_\chi$ or $b_\chi$ follow from the heavy mediator scale and the small kinetic mixing.
This UV completion applies to sufficiently small $a_\chi$ or $b_\chi$ for which the matching can be realized with perturbative couplings, $|\epsilon|\lesssim 1$, and mediator masses well above the invariant masses that dominate freeze-in production.

The same matching also generates $-g_D^2J_D^\mu J_{D\mu}/(2m_{A'}^2)$, which mediates processes such as $\chi_1\chi_1\leftrightarrow\chi_0\chi_0$.
It is important to check that throughout the time evolution, the excited-state abundance is not altered, which takes place if the corresponding rate is comparable to the Hubble rate.
We use a conservative upper bound on $g_D$ by requiring the conversion optical depth $D_{\rm conv}=\int dt\,n_1\langle\sigma v\rangle_{11\to00}<0.1$.
Our benchmarks assume $g_D\lesssim 0.1$ over the reheating-temperature range considered.

The physical elastic-to-inelastic coupling ratio is
\begin{equation}
    \left|\frac{g_{\mathrm{el}}}{g_{\mathrm{tr}}}\right| = \left|\cot(2\theta)\right| = \frac{|f_\eta-f_\xi|}{2|m_D|}
\,.
\end{equation}
Therefore, the interaction is dominantly inelastic whenever
\begin{equation}
    |f_\eta-f_\xi| \,\ll 2|m_D|
\,.
\end{equation}
In particular, in the pseudo-Dirac regime, $|m_D|\gg|f_\eta|,|f_\xi|$, one has $\det\mathcal{M}<0$ and obtains a dominantly inelastic transition CR interaction. 
On the other hand, in the pseudo-Majorana regime, $|\bar f|\gg|m_D|\gg|\delta f|/2$, one has $\det\mathcal{M}>0$, which results in a dominantly inelastic transition AM interaction. 
The sign of $\det\mathcal{M}$ therefore selects a dominantly inelastic AM or CR interaction.
At tree level, integrating out the DP generates no transition dipole or Rayleigh operator.

The dimension-6 transition operators do not mediate the two-body decay $\chi_1\to\chi_0+\gamma$.
Their transverse vertex, proportional to $q^2\,g_{\mu\nu}-q_\mu q_\nu$, vanishes when contracted with an on-shell photon, as does the underlying kinetic-mixing insertion.
On the other hand, for dipole iDM, $\chi_1$ decays semi-visibly into $\chi_0$ and a photon, resulting in strong constraints from displaced decays at beam dumps and forward physics experiments~\cite{Jodlowski:2023ohn,Dienes:2023uve}.

The QED amplitude $\gamma^\ast\to2\gamma$ vanishes by Furry's theorem.
Therefore, at low energies, the signatures of the setup defined by \cref{eq:L_setup} are determined solely by AM or CR operators, while above the EW scale, the effective theory given by \cref{eq:L_int} is promoted to the EW gauge-invariant form, which leads to the three-body decay $\chi_1\to\chi_0 + Z^\ast \to \chi_0 +\nu\bar{\nu}$.
The term $F_{\mu\nu}$ in \cref{eq:L_int} is replaced by the hypercharge tensor $B_{\mu\nu}/\cos\theta_W$, and the resulting iDM couplings to the $Z$ boson are then proportional to the same coefficients as the photon couplings, but with an additional factor of minus the tangent of the weak mixing angle, $g_{Z \chi_0\chi_1} = -\tan\theta_W \,g_{\gamma\chi_0\chi_1}$, where $g_{\gamma\chi_0\chi_1} \in \{ 1/\Lambda_E, 1/\Lambda_M, a_\chi, b_\chi\}$.
A hypercharge completion of the Rayleigh operators similarly generates $\gamma\gamma$, $\gamma Z$, and $ZZ$ interactions.
In large part of the parameter space, the momentum transfer satisfies $q^2\ll m_Z^2$, and the corresponding $Z$-exchange amplitudes are suppressed.
However, for theoretical consistency, we include the $Z$ contribution in the numerical freeze-in and Drell--Yan process calculations, while photon-only expressions are used only in their low-energy domain of validity.

At the momentum transfers relevant to excited-state decay and direct detection, $Z$ exchange is suppressed by the weak scale. 
This suppression does not generally justify neglecting $Z$-mediated cosmological or collider production, which is included in the numerical calculations.
We numerically computed the three-body decay width $\chi_1 \to \chi_0 + \nu\bar{\nu}$, finding that it is strongly suppressed, as described by the following estimate:
\begin{equation}
\label{eq:chi1_to_chi0nunu}
    \Gamma^{\mathrm{AM}/\mathrm{CR}}_{\chi_1 \to \chi_0 +\nu\nu} \simeq 10^{-3} \,\frac{\alpha \,c_\chi^2\,\Delta^9}{m_Z^4}
\,,
\end{equation}
where $c_\chi\in\{a_\chi,b_\chi\}$.
Therefore, the invisible decay is suppressed by the factor $10^{7}\,\alpha^{-3}\,m_e^8\, (m_Z\,\Delta)^{-4}$ compared to the three-photon decay, and for $\Delta\gtrsim 6 (7)$ keV, the three photon decay dominates for AM (CR). 

Once kinematically accessible, decays such as $\chi_1\to\chi_0+e^+e^-$, and eventually hadronic channels, give strong intensity-frontier and collider constraints on AM and CR iDM, including the freeze-out parameter space studied in Ref.~\cite{Jodlowski:2023ohn}.
We instead focus on the complementary regime $\Delta<2m_e$, which is also naturally realized in the pseudo-Dirac limit.
The leading visible decay is then the loop-induced process $\chi_1\to\chi_0+\gamma^\ast\to\chi_0+3\gamma$, whose width and spectrum we calculate below.

\section{Multiphoton decay widths and spectra}
\label{sec:decay}
The excited-state lifetime and photon spectrum determine the X-ray sensitivity to the freeze-in parameter space.
We first derive the three-photon results for AM and CR interactions and compare them with the two-photon Rayleigh decays obtained in the Appendix~\ref{app:Ray}.

\subsection{Total decay widths}
\label{sec:Gamma}
\begin{figure}[tb]
    \centering
    \includegraphics[width=0.45\textwidth]{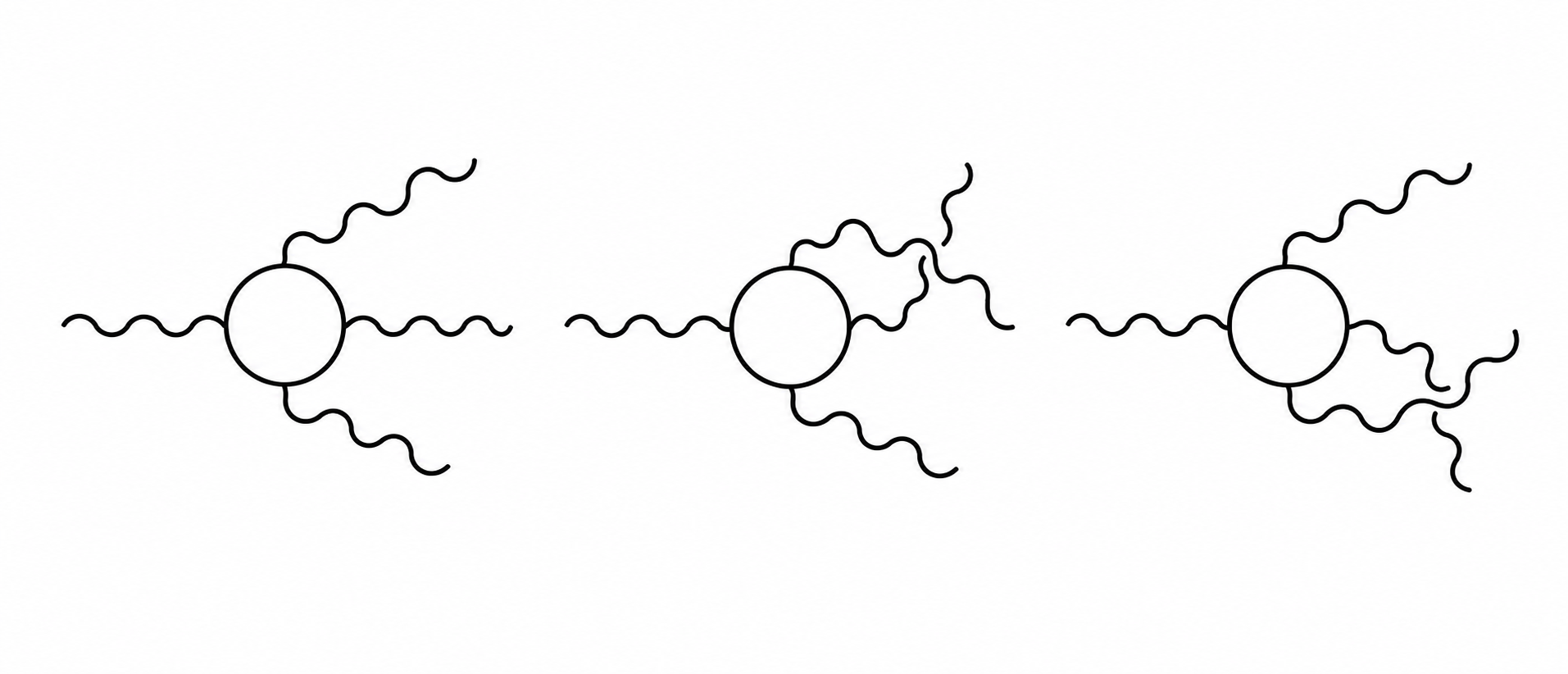}
    \caption{
        One-loop diagrams for $\gamma^\ast\to3\gamma$~\cite{Heisenberg:1936nmg,McDermott:2017qcg}.
        Both orientations of the charged-fermion loop are included.
        }
    \label{fig:Feynman_diagrams}
\end{figure}

For a decay into $n$ photons, with $n=2$ or $3$, the width is
\begin{equation}
\begin{aligned}
    \label{eq:chi1_23gamma}
    \Gamma_{\chi_1\to\chi_0 +(2)3\gamma}& = \frac{1}{(2!)3!}\frac{1}{4 m_{\chi_1}}\int d\Phi_{(3)4} \left|{\cal M}\right|_{\chi_1\to\chi_0 +(2)3\gamma}^2 
\,,
\end{aligned}
\end{equation}
where the squared matrix element is averaged over the initial spin and summed over final spins and polarizations; the factor $1/n!$ accounts for the identical photons.
For AM and CR interactions, the three-photon channel proceeds through an off-shell photon of momentum $k$, followed by $\gamma^\ast(k)\to3\gamma$.
To leading order in $\Delta/m_{\chi_0}$, the spin-averaged squared matrix elements take the form
\begin{align}
    \overline{|\mathcal M|^2}_{\chi_1\to\chi_0 +3\gamma \,\,\rm AM} &\simeq \left[ \frac{8 a_\chi^2m_{\chi_0}^2}{k^2} \left(\Delta^2+2k^2\right) \right] \overline{|\mathcal M|^2}_{\gamma^\ast\to3\gamma}
    \,,
    \nonumber \\
    \overline{|\mathcal M|^2}_{\chi_1\to\chi_0 +3\gamma \,\,\rm CR} &\simeq \left[ \frac{8 b_\chi^2m_{\chi_0}^2}{k^2} \left(\Delta^2-k^2\right) \right] \overline{|\mathcal M|^2}_{\gamma^\ast\to3\gamma}
\,,
    \label{eq:M2}
\end{align}
where $\overline{|\mathcal M|^2}_{\gamma^\ast\to3\gamma}$ term corresponds to the  1-loop box diagrams with electrons (the dominant contribution among charged SM fermions) on internal lines, and was calculated in the  $\sqrt{k^2} \ll 2m_e$ limit first by Euler and Heisenberg~\cite{Heisenberg:1936nmg}.
Their result for the virtual-photon width is
\begin{equation}
    \label{eq:gamma_star_width}
    \Gamma^0_{\gamma^\ast \to 3\gamma}(\sqrt{k^2}) =  \frac{17\alpha^4 k^{9}}{2^7 \times 3^6 \times 5^3 \times \pi^3 m_e^8}
\,,
\end{equation}
where the superscript $0$ denotes the EH limit.
For the general case, we implement the amplitudes shown in  \cref{fig:Feynman_diagrams} in Package-X~\cite{Patel:2015tea} (there are two fermion flow directions -- clockwise and anticlockwise, equal due to the CP symmetry of QED), which allows convenient numerical evaluation of the resulting Passarino--Veltman functions.

\begin{figure}[t]
    \centering
    \includegraphics[width=0.48\textwidth]{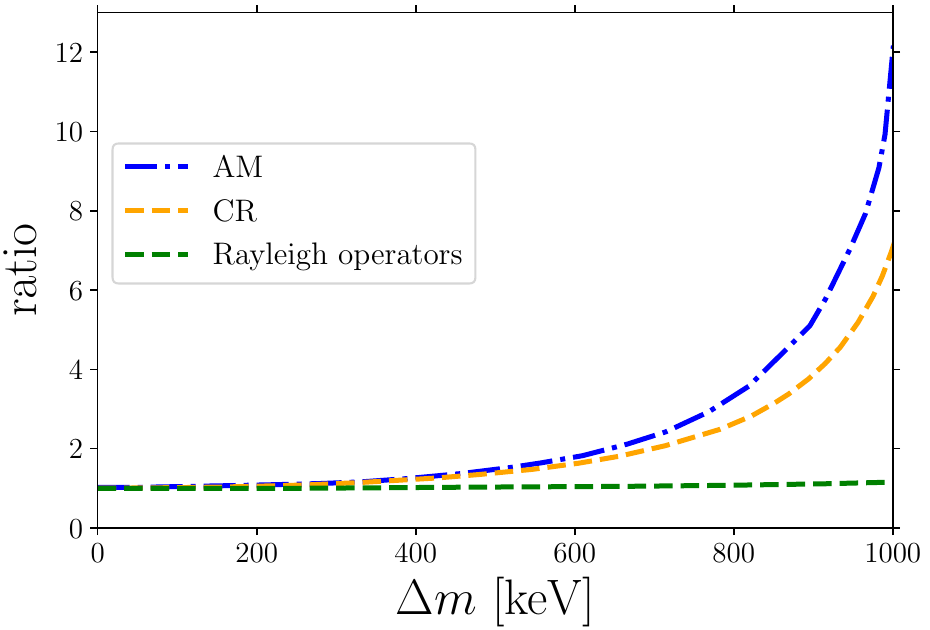}
    \caption{
        Ratio of the full one-loop three-photon width to the EH result as a function of $\Delta$ for AM (blue) and CR (orange).
        The AM virtuality distribution places more weight at large $k^2$, where departures from the EH limit are larger.
        The green curve shows the lepton-mass correction to the scalar-mediated two-photon widths in the Higgs-portal realization, using the low-energy matching described in the Appendix~\ref{app:Ray}.
        }
    \label{fig:beyond_HE}
\end{figure}

At leading order in $\Delta/m_{\chi_0}$, integration over the remaining phase-space variables reduces the width to a one-dimensional integral over the photon virtuality $k^2$~\cite{Pospelov:2008jk,Krnjaic:2025zjl}.
We obtain
\begin{equation}
\begin{aligned}
    \Gamma^{\mathrm{AM,\,0}}_{\chi_1\to\chi_0 +3\gamma} &\simeq \frac{a_{\chi}^2}{2\pi^2} \int_0^{\Delta^2}  \frac{dk^2}{\sqrt{k^2}}\left(\Delta^2-k^2\right)^{1/2} 
    \\
    & \quad\quad\quad \quad\quad \quad   \left(\Delta^2+2k^2\right) \Gamma_{\gamma^\ast \to 3\gamma}(\sqrt{k^2}) 
\,,
\\
    \Gamma^{\mathrm{CR,\,0}}_{\chi_1\to\chi_0 +3\gamma} &\simeq \frac{b_{\chi}^2}{2\pi^2} \int_0^{\Delta^2}  \frac{dk^2}{\sqrt{k^2}}\left(\Delta^2-k^2\right)^{3/2} \Gamma_{\gamma^\ast \to 3\gamma}(\sqrt{k^2}) 
\,.
    \label{eq:widths}
\end{aligned}
\end{equation}
Inserting the EH expression in \cref{eq:gamma_star_width} gives
\be
    \label{eq:Gamma_an}
    \Gamma^0_{\rm AM} &= \frac{17\alpha^4 a_\chi^2} {3^7\times 5^4\times 7\times 13\times \pi^5} \frac{\Delta^{13}}{m_e^8}
\,,
\\
    \Gamma^0_{\rm CR} &= \frac{17\alpha^4 b_\chi^2} {3^7\times 5^4 \times 7 \times 11 \times 13 \times \pi^5} \frac{\Delta^{13}}{m_e^8}
\,.
\ee
Numerically, these widths are
\be
    \label{eq:Gamma_num}
    \Gamma^0_{\rm AM} &\simeq 1.05\times10^{-23}\,{\rm s}^{-1} \left(\frac{a_\chi}{10^{-8}\GeV^{-2}}\right)^2 \left(\frac{\Delta}{900\keV}\right)^{13}
\,,
\\
    \Gamma^0_{\rm CR} &\simeq 9.56\times10^{-25}\,{\rm s}^{-1} \left(\frac{b_\chi}{10^{-8}\GeV^{-2}}\right)^2 \left(\frac{\Delta}{900\keV}\right)^{13}
\,.
\ee
The reference splitting $\Delta=900\keV$ lies below, but close to, the dielectron threshold $2m_e\simeq1022\keV$.
It gives a cosmologically long-lived excited state for the reference couplings and strong INTEGRAL/SPI sensitivity, as discussed in \cref{sec:xray_recast}.
The steep $\Delta^{13}$ dependence follows from the low-energy $\gamma^\ast\to3\gamma$ amplitude together with the four-body phase space.
At splittings close to $2m_e$, higher terms in the low-energy expansion become important.
We include them by evaluating the full electron-loop amplitude.
The resulting ratios of the total widths to the EH expressions are shown in \cref{fig:beyond_HE}.

\subsection{Photon spectrum in the excited-state rest frame}
\label{sec:gamma_spectrum}
We obtain the photon-number spectrum in the $\chi_1$ rest frame by convolving the virtual-photon decay spectrum with the virtuality distribution in $\chi_1\to\chi_0+\gamma^\ast$.

In the EH limit, the single-photon spectrum in the $\gamma^\ast$ rest frame is~\cite{Pospelov:2008jk}
\be
  \frac{dN_\ast}{dz} &= \frac{1}{17}\,z^3\!\left(1715 - 3105 z + \frac{2919}{2} z^2\right)
\,,
\ee
where $z = 2E_\gamma^\ast/\sqrt{k^2} \in (0,1)$, and is normalized such that $\int_0^1 dz\, dN_\ast/dz = 3$, while $E_\gamma^\ast$ is the photon energy in the $\gamma^\ast$ rest frame.
We define the dimensionless photon energy in the $\chi_1$ rest frame as
\begin{equation}
  x = \frac{E_\gamma}{\Delta}\,, \qquad 0 < x < 1
\,.
\end{equation}
We also introduce the dimensionless virtuality
\begin{equation}
    y = \frac{k^2}{\Delta^2} \in (0,1)\,, \qquad \beta(y) = \sqrt{1-y}\,,
\end{equation}
so that in the $\chi_1$ rest frame the off-shell photon has energy $k^0 \simeq \Delta$ and three-momentum $|\mathbf{k}| = \Delta\,\beta(y)$ to LO in $\Delta/m_{\chi_0}$.
Following~\cite{Pospelov:2008jk,Krnjaic:2025zjl,Jeesun:2026ryo}, we neglect correlations between the polarization of the intermediate virtual photon and its decay angles. 
Then, at fixed $y$, we isotropically boost the spin-averaged $\gamma^*\to3\gamma$ spectrum to the $\chi_1$ rest frame, obtaining
\begin{equation}
    \left.\frac{dN}{dx}\right|_{y} =  \frac{1}{\beta(y)} \int_{z_{-}(x,y)}^{z_{+}(x,y)} dz\, \frac{1}{z}\,\frac{dN_\ast}{dz}(z)
\,,
\end{equation}
with kinematic limits
\begin{equation}
    z_{-}(x,y) = \frac{2x}{1+\beta(y)}\,, \quad  z_{+}(x,y) = \min\!\left(1,\frac{2x}{1-\beta(y)}\right)
\,.
\end{equation}
The fixed-$y$ spectrum vanishes when $z \to z_+$.
Convolving with the virtuality distribution then gives
\begin{equation}
  \frac{dN}{dx} = \int_0^1 dy\, p(y)\, \underbrace{\frac{1}{\beta(y)} \int_{z_{-}(x,y)}^{z_{+}(x,y)} dz\, \frac{1}{z}\,\frac{dN_\ast}{dz}(z)}_{\text{boosted }\gamma^\ast\to 3\gamma \text{ spectrum at fixed }y}
\!,
\end{equation}
where
\begin{equation}
  p(y) = \frac{1}{\Gamma}\,\frac{d\Gamma}{dy}
\end{equation}
is the normalized probability density for the virtuality
 $y = k^2/\Delta^2$, obtained from the squared matrix element
for $\chi_1 \to \chi_0 \gamma^\ast$ and the four-body phase space.
By construction, $\int_0^1 dy\,p(y)=1$.

Combining \cref{eq:M2,eq:gamma_star_width} gives the EH virtuality distributions
\be
    p_{\rm AM}(y) &= \frac{15015}{2816}\, y^4(1+2y)\sqrt{1-y}
    \,,
    \nonumber\\
    p_{\rm CR}(y) &=\frac{15015}{256} y^4(1-y)^{3/2} 
\,.
\ee
The resulting AM photon-number spectrum is
\be
    \left.\frac{dN}{dx}\right|_{\rm AM} = \,\frac{13}{1530}\,x^3\left[(1-x)\,g(x) - 11033820\,x^2\log x\right]
\,,
\ee
where
\begin{align}
    g(x) &= 550900 x^8-3698900 x^7+10820836 x^6 \nonumber\\
    &-18002189 x^5+18229111 x^4-9780689 x^3 \nonumber\\
    &-3749489 x^2-6055100 x+651700
\,.
\end{align}
For CR, we recover the spectrum of the vector-mediator model studied in Ref.~\cite{Krnjaic:2025zjl},
\be
    \left.\frac{dN}{dx}\right|_{\rm CR}  = \frac{143}{1530} x^3 \Bigg[(1-x)\, f(x) -11033820\, x^2 \log x\Bigg]
\,,
\ee
where
\begin{align}
    f(x) &= -275450 x^8 +1849450 x^7 -5630414 x^6  \nonumber\\
    &+10465561 x^5 -13688639 x^4 +14321161 x^3 \nonumber\\
    & -15834839 \,x^2 -2377850 \,x +137200
\,.
\end{align}

\begin{figure}[t]
    \centering
    \includegraphics[width=0.48\textwidth]{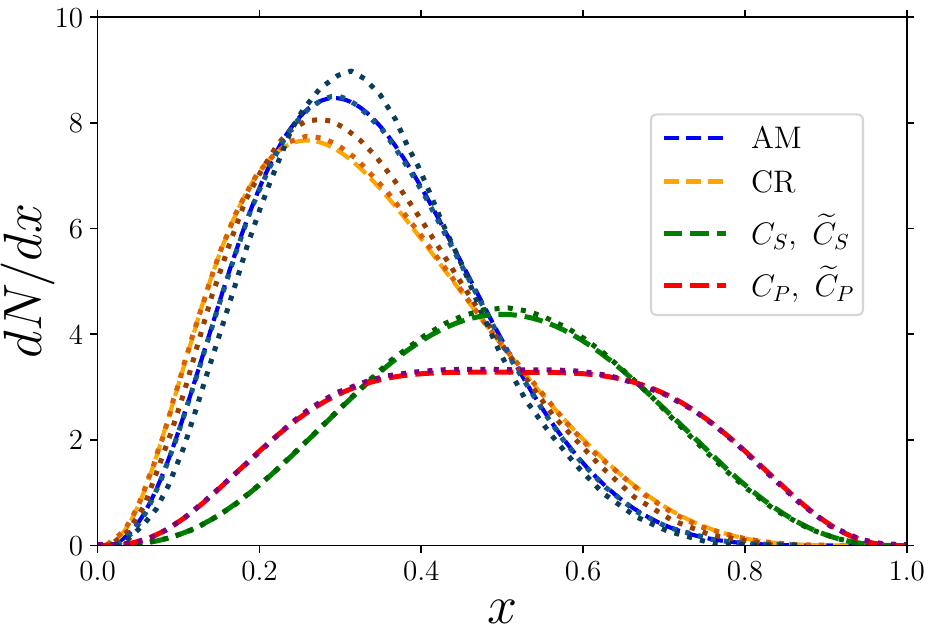}
    \caption{
            Normalized single-photon spectra from $\chi_1\to\chi_0 +3\gamma$ for AM and CR operators.
            The thick dashed and solid curves show the analytical results, while the dotted curves include the full electron-loop mass dependence.
            Darker and darkest shades correspond to $\Delta=500\keV$ and $900\keV$, respectively.
            The AM spectrum peaks at larger $x$ than the CR spectrum, consistent with its larger weight at high photon virtuality and hence smaller boost of the three-photon system.
            We also show the spectra for the transition Rayleigh operators. The details are discussed in the Appendix~\ref{sec:xray_constr}.
            }
    \label{fig:spectra}
\end{figure}

Figure~\ref{fig:spectra} compares the EH spectra with those obtained from the full one-loop amplitudes, together with the Rayleigh spectra derived in the Appendix~\ref{app:Ray}.
Each spectrum is normalized to the photon multiplicity, so the figure isolates changes in spectral shape.
The total-width corrections are shown separately in \cref{fig:beyond_HE}.
The larger finite-lepton-mass effects in the vector-mediated three-photon channel are consistent with the $S$-wave electron--positron threshold of the vector current, compared with the $P$-wave threshold of a CP-even scalar.
In the Higgs-portal scalar amplitude, mass-proportional Yukawa couplings also give comparable low-energy electron, muon, and tau loop contributions, diluting the relative effect of the electron threshold.

\section{DM relic density from UV-dominated freeze-in}
\label{sec:freezein}
With a heavy mediator, dimension-6 interactions make freeze-in sensitive to the highest temperatures attained by the thermal bath, which is in contrast to the infrared-dominated production characteristic of a light dark photon.

In the photon-dominated regime, the leading production process is SM charged-fermion annihilation, $f\overline f\to\gamma^*\to\chi_0\chi_1$; our numerical calculation also includes the correlated $Z$ contribution and its interference with photon exchange.
We also note that freeze-out through these operators was studied in Ref.~\cite{Jodlowski:2023ohn}, where the corresponding benchmarks were found to be excluded by existing observations.

We assume instantaneous reheating to a temperature $T_{\mathrm{RH}}$, followed by radiation domination and adiabatic expansion, with a vanishing initial dark-sector abundance.
While taking into account the dynamics of the reheating is possible within \texttt{micrOMEGAs}~\cite{Belanger:2014vza,Belanger:2018ccd}, it requires making additional physical assumptions and introducing more free parameters, thus, being outside of our scope.

Under these assumptions, the coupled number densities obey the following Boltzmann equations:
\begin{align}
    \dot n_0+3Hn_0 &=\gamma_{\mathrm{pair}}\bigl(1-r_0r_1\bigr) +\mathcal C_{1\leftrightarrow0} \,,
    \nonumber\\
    \dot n_1+3Hn_1 &=\gamma_{\mathrm{pair}}\bigl(1-r_0r_1\bigr) -\mathcal C_{1\leftrightarrow0}\,,
\label{eq:fi-coupled-boltzmann}
\end{align}
where $r_i= n_i/n_i^{\mathrm{eq}}$ for $i=0,1$, $\gamma_{\mathrm{pair}}$ is the sum of the equilibrium reaction densities for all bath processes producing $\chi_0$-$\chi_1$, while $\mathcal C_{1\leftrightarrow0}$ describes scatterings and dark-sector reactions that redistribute the two populations.  
Excited-state decays are negligible during production because $\Gamma_{\chi_1}\ll H$.
Near the intersections with the INTEGRAL/SPI limits, $\tau_{\chi_1}\gg13.8\,\mathrm{Gyr}$, so subsequent decay depletion is also negligible.
Adding the two equations eliminates the conversion terms. 
Neglecting inverse annihilations in the freeze-in regime, the total yield $Y_\chi = Y_0 + Y_1$ obeys
\begin{equation}
    \frac{\mathrm dY_\chi}{\mathrm dT} =-\frac{2\mathcal A_s(T)}{sHT}\, \gamma_{\mathrm{pair}}(T)
    \,,
    \quad 
    \mathcal A_s(T) = 1+\frac{T}{3g_{\ast s}}\frac{\mathrm dg_{\ast s}}{\mathrm dT}
\,,
    \label{eq:fi-total-yield}
\end{equation}
where $s$, $\gamma_{\mathrm{pair}}(T)$, and $g_{\ast s}$ are the entropy density of the thermal bath, the total reaction density for DS production, and the effective number of relativistic degrees of freedom contributing to the entropy density, respectively.
If conversion is inefficient, pair production gives $Y_0=Y_1$ and hence a present excited-state fraction $f_1\simeq1/2$ in the small-splitting, long-lifetime limit.
The present abundance is
\begin{align}
    \Omega_\chi h^2 &=\frac{s_0h^2}{\rho_c} \left(m_{\chi_0}Y_0+m_{\chi_1}Y_1\right)
    \nonumber\\
    &\simeq 2.742\times10^8 \left(\frac{m_{\chi_0}}{\mathrm{GeV}}\right) Y_\chi
    \,,
\label{eq:fi-relic-density}
\end{align}
where $s_0$, $h$, and $\rho_c$ are the present-day entropy density, the reduced Hubble parameter defined by $H_0 = 100\,h\,\mathrm{km \,s^{-1}\,Mpc^{-1}}$, and the present-day critical density of the Universe, respectively; the second line holds for $\Delta\ll m_{\chi_0}$, which is assumed throughout.

For reference, in the Maxwell--Boltzmann approximation the reaction density for $f\overline f\to\chi_0\chi_1$ is
\begin{equation}
    \gamma_f(T) =\frac{g_f^2T}{32\pi^4} \int_{s_{\min,f}}^{\infty} \mathrm ds\, \frac{\lambda(s,m_f^2,m_f^2)}{\sqrt{s}}\, \sigma_f(s)\,K_1\!\left(\frac{\sqrt{s}}{T}\right)
\,,
\label{eq:fi-reaction-density}
\end{equation}
where $K_1$ is the modified Bessel function of the second kind, $\lambda(x,y,z)=(x-y-z)^2-4yz$, $s_{\min,f}=\max[4m_f^2,(m_{\chi_0}+m_{\chi_1})^2]$, and $\sigma_f$ is averaged over the initial spin and color degrees of freedom included in $g_f$.  
For degenerate dark states, $m_{\chi_0}=m_{\chi_1}=m_\chi$, and massless bath fermions, the photon-exchange cross sections are~\cite{Chu:2018qrm}
\begin{align}
\label{eq:fi-cross-sections}
    \sigma_f^{\mathrm{AM}}(s) &=\frac{\alpha Q_f^2a_\chi^2}{3N_c^f}\, s\,\beta_\chi^3
    \,,
\nonumber\\
    \sigma_f^{\mathrm{CR}}(s) &=\frac{\alpha Q_f^2b_\chi^2}{3N_c^f}\, \left(s+2m_\chi^2\right)\beta_\chi
    \,,
\end{align}
where $\beta_\chi =\sqrt{1-\frac{4m_\chi^2}{s}}$ and the factor $1/N_c^f$ applies to a color-averaged partonic cross section.  After combining it with the bath multiplicities, each charged fermion carries the weight $N_c^fQ_f^2$.
Cross sections for unequal dark-state masses are given in Ref.~\cite{Jodlowski:2023ohn}.

\subsection{Analytical limits}
Before discussing the results originating from numerical solution to \cref{eq:fi-coupled-boltzmann} obtained with \texttt{micrOMEGAs}~\cite{Belanger:2014vza,Belanger:2018ccd}, we discuss the yield dependence in two opposite limits, $T_{\mathrm{RH}}\ll m_{\chi_0}$ and $T_{\mathrm{RH}}\gg m_{\chi_0}$.
We derive both limits for photon exchange and compare them with the numerical calculation below.

\paragraph{Relativistic production: $T_{\mathrm{RH}}\gg m_{\chi_0}$.}
For $s\gg m_{\chi_i}^2$, the two operators give the same cross section,
\begin{equation}
    \sigma_f(s) \simeq \frac{\alpha Q_f^2c_\chi^2}{3N_c^f}\,s
    \,,
    \qquad
    c_\chi=a_\chi\ \text{or}\ b_\chi
\,,
    \label{eq:fi-rel-cross-section}
\end{equation}
while for Maxwell--Boltzmann distributions the thermal average is $\langle\sigma v\rangle_f= 8\alpha Q_f^2c_\chi^2T^2/N_c^f$.
Taking $g_\ast$, $g_{\ast s}$, and the set of relativistic charged species to be constant over the production interval results in
\begin{equation}
    Y_\chi^{\mathrm{rel}} \simeq \frac{480\sqrt{90}}{\pi^7} \frac{\alpha\,\mathcal Q_{\mathrm{RH}}} {g_{\ast s}(T_{\mathrm{RH}}) \sqrt{g_\ast(T_{\mathrm{RH}})}}\, \overline M_{\mathrm{Pl}} \,c_\chi^2 \,T_{\mathrm{RH}}^3
    \,,
\label{eq:fi-rel-yield}
\end{equation}
where $\overline M_{\mathrm{Pl}}$ is the reduced Planck mass and $\mathcal Q_{\mathrm{RH}} = \sum_{m_f\ll T_{\mathrm{RH}}}N_c^fQ_f^2$.
Combining this result with~\eqref{eq:fi-relic-density} gives the characteristic target scaling
\begin{equation}
    a_\chi^{\mathrm{FI}},\ b_\chi^{\mathrm{FI}} \propto m_{\chi_0}^{-1/2}\, T_{\mathrm{RH}}^{-3/2}
\,.
    \label{eq:fi-rel-scaling}
\end{equation}
Finite masses, changes in the relativistic degrees of freedom, and quantum statistics modify the normalization and introduce threshold structure.
Away from such thresholds, the leading relativistic scaling remains a useful guide.

\paragraph{Boltzmann-suppressed production:
$T_{\mathrm{RH}}\ll m_{\chi_0}$.} In this limit the integral is localized near $T_{\mathrm{RH}}$ and $\sqrt{s}\simeq m_{\chi_0}+m_{\chi_1}$.  
For simplicity, let us temporarily assume $m_{\chi_0}=m_{\chi_1}=m_\chi$,  consider the inverse rates, and invoke the detailed balance condition.
Summing over kinematically accessible charged fermions gives
\begin{align}
\label{eq:fi-nr-annihilation}
    \sum_f\langle\sigma v_{\mathrm{rel}}\rangle_{ \chi_0\chi_1\to f\overline f}^{\mathrm{AM}} &\simeq  4\alpha\,\mathcal Q_m a_\chi^2m_\chi T
      \,,
    \nonumber\\
    \sum_f\langle\sigma v_{\mathrm{rel}}\rangle_{ \chi_0\chi_1\to f\overline f}^{\mathrm{CR}} &\simeq 4\alpha\,\mathcal Q_m b_\chi^2m_\chi^2
      \,.
\end{align}
The AM channel is $p$-wave, whereas the CR channel is $s$-wave.  
The charge sum $\mathcal Q_m$ includes fermions open at the scale $\sqrt{s}\simeq2m_\chi$, rather than only species that are relativistic at $T_{\mathrm{RH}}$.
Using $n_i^{\mathrm{eq}}\simeq g_\chi(m_\chi T/2\pi)^{3/2} e^{-m_\chi/T}$ with $g_\chi=2$, and keeping the leading threshold cross sections in~\eqref{eq:fi-nr-annihilation}, the temperature integral can be done analytically.  
We find
\begin{align}
\label{eq:fi-nr-yields}
    Y_\chi^{\mathrm{AM}} &\simeq 2\mathcal K_{\mathrm{NR}}\,a_\chi^2 \,m_\chi^3 \,e^{-2x_{\mathrm{RH}}} \left[1+\mathcal O\!\left(x_{\mathrm{RH}}^{-1} \right)\right]
    \,,
    \\
    Y_\chi^{\mathrm{CR}} &\simeq \mathcal K_{\mathrm{NR}}\,b_\chi^2 \,m_\chi^3 \left(1+2x_{\mathrm{RH}}\right)\, e^{-2x_{\mathrm{RH}}} \left[1+\mathcal O\!\left(x_{\mathrm{RH}}^{-1} \right)\right]
    \nonumber
    \,.
\end{align}
We defined $x_{\mathrm{RH}}= m_\chi/T_{\mathrm{RH}}$ and
\begin{equation}
    \mathcal K_{\mathrm{NR}} = \frac{45\sqrt{90}}{8\pi^6} \frac{g_\chi^2\alpha\,\mathcal Q_m\overline M_{\mathrm{Pl}}} {g_{\ast s}(T_{\mathrm{RH}}) \sqrt{g_\ast(T_{\mathrm{RH}})}}
\,.
    \label{eq:fi-knr}
\end{equation}
For a nonzero splitting, the common exponential is replaced by $\exp[-(m_{\chi_0}+m_{\chi_1})/T_{\mathrm{RH}}]$, with relative corrections of order $\Delta/m_{\chi_0}$ in the prefactors.
In the asymptotic limit,~\eqref{eq:fi-nr-yields} implies
\begin{align}
    a_\chi^{\mathrm{FI}} &\propto m_\chi^{-2}\,e^{m_\chi/T_{\mathrm{RH}}}
    \,,
    \nonumber\\
    b_\chi^{\mathrm{FI}} &\propto T_{\mathrm{RH}}^{1/2}\,m_\chi^{-5/2}\, e^{m_\chi/T_{\mathrm{RH}}}
    \,,
    \nonumber\\
    \frac{a_\chi^{\mathrm{FI}}}{b_\chi^{\mathrm{FI}}} &\simeq \sqrt{\frac{m_\chi}{T_{\mathrm{RH}}}}
    \,,
    \label{eq:fi-nr-scaling}
\end{align}
where the last relation is a consequence of the $p$-wave suppression of AM production. 

\subsection{Numerical results and comparison with the analytical limits}
We implement the AM and CR interactions in \texttt{FeynRules}~\cite{Alloul:2013bka} and solve the coupled freeze-in equations with \texttt{micrOMEGAs}~\cite{Belanger:2014vza,Belanger:2018ccd}.
The calculation retains the exact dark-state masses, the full $2\to2$ matrix elements, thermal distributions, the temperature-dependent SM equation of state, and all kinematically available annihilation channels included in the model.

We consider three illustrative benchmarks for the reheating temperature, 
$T_{\mathrm{RH}}=0.1$, $1$, and $10\,\mathrm{GeV}$.
In \cref{fig:main_results}, we show the resulting relic density contours for these temperatures in black, red, and blue, respectively.
For each benchmark, we verify that the dark states remain out of chemical equilibrium.
While lower reheating temperatures satisfying $T_{\mathrm{RH}}\gtrsim 6$ MeV are possible~\cite{Barbieri:2025moq}, their contours can be obtained by rescaling (neglecting small effects associated with changes in the thermal degrees of freedom), as discussed below.

In the photon-dominated regime, the contours approach the asymptotic scaling described above: for $m_{\chi_0}\ll T_{\mathrm{RH}}$, the AM and CR contours coincide and decrease as $m_{\chi_0}^{-1/2}$.  For $m_{\chi_0}\gg T_{\mathrm{RH}}$, both contours turn up exponentially, while the AM target lies above the CR target by the factor $\simeq\sqrt{m_{\chi_0}/T_{\mathrm{RH}}}$.  
The broad minimum at $m_{\chi_0}/T_{\mathrm{RH}}=\mathcal O(1)$ is the interpolation region in which the full numerical result is required.

To validate the analytical approximations, we compare them with a numerical benchmark containing photon exchange only.
For example, for a relativistic benchmark of $T_{\mathrm{RH}}=10\,\mathrm{GeV}$ and $m_{\chi_0}=0.1\,\mathrm{GeV}$,~\cref{eq:fi-relic-density,eq:fi-rel-yield} give $a_\chi^{\mathrm{FI}}=b_\chi^{\mathrm{FI}} \simeq 1.40\times10^{-13}\,\mathrm{GeV}^{-2}$, while the numerical photon-only calculation gives $a_\chi^{\mathrm{FI}}=b_\chi^{\mathrm{FI}} \simeq 1.42\times10^{-13}\,\mathrm{GeV}^{-2}$, a difference of about one percent.
As a Boltzmann-suppressed benchmark, we take $T_{\mathrm{RH}}=1\,\mathrm{GeV}$ and $m_{\chi_0}=20\,\mathrm{GeV}$, for which $g_\ast=g_{\ast s}=61.75$, $\mathcal Q_m=20/3$, and~\eqref{eq:fi-nr-yields} gives $a_\chi^{\mathrm{FI}} \simeq 2.44\times10^{-6}\,\mathrm{GeV}^{-2}$, $b_\chi^{\mathrm{FI}} \simeq 5.38\times10^{-7}\,\mathrm{GeV}^{-2}$.
The corresponding numerical values are $a_\chi^{\mathrm{FI}} \simeq 2.41\times10^{-6}\,\mathrm{GeV}^{-2}$ and $b_\chi^{\mathrm{FI}} \simeq 5.48\times10^{-7}\,\mathrm{GeV}^{-2}$, respectively.  
For these benchmarks, the analytical approximations reproduce the numerical photon-only calculation at the percent level. 
The hypercharge contours in \cref{fig:main_results} include the correlated $Z$ interaction and its interference with photon exchange. For $T_{\mathrm{RH}}=0.1$ and $1\,\mathrm{GeV}$, they remain close to the photon-only contours over the phenomenologically relevant branches, where production predominantly samples invariant masses below the weak scale and the $Z$ pole is strongly Boltzmann suppressed. For $T_{\mathrm{RH}}=10\,\mathrm{GeV}$, the thermal tail can instead sample the $Z$ pole when $m_{\chi_0}+m_{\chi_1}<m_Z$, enhancing production and lowering the required Wilson coefficients. The off-shell correction is controlled by the invariant masses sampled in production, which approach $\sqrt{s}\simeq m_{\chi_0}+m_{\chi_1}$ on the Boltzmann-suppressed branch, rather than by $T_{\mathrm{RH}}$ alone.

The numerical contours must be interpreted within the freeze-in and effective field theory (EFT) domains.
Neither asymptotic expansion is controlled near $m_{\chi_0}/T_{\mathrm{RH}}\sim1$.
Moreover, for $m_{A'}=1\,\mathrm{TeV}$ and the assumed small dark gauge coupling, the large Wilson coefficients in the Boltzmann-suppressed comparison benchmark cannot be matched with perturbative kinetic mixing.
That benchmark therefore tests the asymptotic calculation rather than defining an allowed UV completion.
Finally, production on this branch occurs near $\sqrt{s}\simeq m_{\chi_0}+m_{\chi_1}$, even when $T_{\mathrm{RH}}$ is small.
The contact description requires $\max(T_{\mathrm{RH}},m_{\chi_0}+m_{\chi_1})\ll m_{A'}$, together with perturbative matching coefficients.

\section{Experimental constraints and projected sensitivities}
\label{sec:Bounds}

\subsection{X-rays constraints}
\label{sec:xray_recast}

The decays $\chi_1\to\chi_0+3\gamma$ and $\chi_1\to\chi_0+2\gamma$ produce continuous photon signals with energies $0<\omega<\Delta$.  
At LO in $\Delta/m_{\chi_0}$, the differential photon flux integrated over a region of interest (ROI) is
\begin{equation}
    \frac{d\Phi_X}{d\omega} =\frac{f_1\,D}{4\pi m_{\chi_0}}\, \Gamma_X\,\frac{1}{\Delta}\, \frac{dN_X \!\left(\frac{\omega}{\Delta}\right)}{dx}
    \,,
    \label{eq:xray_flux}
\end{equation}
where $X\in\{\mathrm{AM},\mathrm{CR},S,P\}$, $D=\int_{\rm ROI}d\Omega\int_{\rm l.o.s.}d\ell\,\rho_{\rm DM}$ is the decay $D$-factor, and $f_1=\Omega_{\chi_1}/\Omega_{\rm DM}$ is the present-day excited-state fraction.  
The widths $\Gamma_X$ in \cref{eq:xray_flux} are the $2$ or $3\gamma$ widths, given by \cref{eq:G_Rayleigh} and \cref{eq:widths}, respectively.
Symmetric pair production and negligible late-time depletion give $f_1\simeq1/2$ in the long-lived regime.
We use this value for the numerical limits quoted below but the evolved excited-state fraction inferred from \texttt{micrOMEGAs} for the combined results shown in \cref{fig:main_results}.

We use the INTEGRAL/SPI analysis of Ref.~\cite{Krnjaic:2025zjl}, based on 16 years of data over $30\,\mathrm{keV}<\omega<8\,\mathrm{MeV}$ and the ROI $-47.5^\circ<\ell,b<47.5^\circ$.
That analysis assumes an NFW halo profile with $D=0.9\times10^{23}\,\mathrm{GeV}\,\mathrm{cm}^{-2}$ and fits the signal jointly with unresolved point sources, inverse-Compton emission, positronium emission, and nuclear lines.

The resulting 95\% credible lower limit on the lifetime--mass product is shown in the left panel of Fig.\,5 therein.
We denote it as
\begin{equation}
    L(\Delta) = \tau_{\chi_1}m_{\chi_0}
    \,.
    \label{eq:LV95}
\end{equation}

For CR iDM this limit applies directly, since the CR matrix element in \cref{eq:M2} has the same virtuality kernel, proportional to $(\Delta^2-k^2)/k^2$, as the vector-mediator model of Ref.~\cite{Krnjaic:2025zjl}, after the replacement $\epsilon g/m_{A'}^2\to b_\chi$.  
Consequently, both the normalized spectrum, including the full one-loop corrections to $\gamma^*\to3\gamma$, and the decay morphology are identical.  
We therefore identify
\begin{equation}
    L_{\rm CR}(\Delta)=L(\Delta),
    \qquad
    \frac{m_{\chi_0}}{\Gamma_{\rm CR}}>2f_1L_{\rm CR}(\Delta).
    \label{eq:CR_lifetime_bound}
\end{equation}
At $\Delta=900\,\mathrm{keV}$, the published limit is~\cite{Krnjaic:2025zjl}
\begin{equation}
    L_{\rm CR}(900\,\mathrm{keV})
    \simeq 7.36\times10^{23}\,\mathrm{GeV}\,\mathrm{s}.
    \label{eq:CR_limit_900}
\end{equation}

On the other hand, the AM spectrum is not identical to the vector template, so an exact limit would require repeating the likelihood analysis.  
In a linearized Gaussian treatment, the corresponding sensitivity recast is
\begin{equation}
    \frac{L_{\rm AM}}{L_{\rm CR}} \simeq \left[ \frac{(t^{\rm AM})^TW_\perp t^{\rm AM}} {(t^{\rm CR})^TW_\perp t^{\rm CR}} \right]^{1/2}
    \,,
    \label{eq:AM_CR_fisher_recast}
\end{equation}
where $t^X$ is the response-convolved unit-amplitude template and $W_\perp$ is the inverse covariance after marginalizing the background nuisance directions.  
These quantities are not provided with the published limit in~\cite{Krnjaic:2025zjl}.  
The normalized spectra are nevertheless similar, as shown in \cref{fig:spectra}. 
To quantify the similarity before response weighting, we use a measure of alignment of the shapes, the unweighted shape cosine, which is defined as the normalized $L^2$ overlap,
\begin{equation}
    \cos(X,Y) = \frac{\int_0^1dx\,s_X(x)s_Y(x)} {\left[\int_0^1dx\,s_X^2(x)\int_0^1dx\,s_Y^2(x)\right]^{1/2}}
\,,
    \label{eq:flat_shape_cosine}
\end{equation}
where $s_X(x)=dN_X/dx$ and $X$ indicates the operator.
In the present case, $X=\,$AM/CR, while Rayleigh operators are discussed in the Appendix~\ref{sec:xray_constr}.
The shape cosine measures the similarity of the spectra, while their relative normalization in the quadratic sensitivity estimate is described by
\begin{equation}
    \label{eq:normaliza}
    \mathcal R_{X/Y}^{\mathcal B} \equiv \left[ \frac{\displaystyle\int_{\mathcal B}dx\,s_X^2(x)} {\displaystyle\int_{\mathcal B}dx\,s_Y^2(x)} \right]^{1/2}
    \,,
\end{equation}
where $s_X(x)=dN_X/dx$ retains its full photon-number normalization and $\mathcal B$ denotes the integration interval. 
With uniform weighting, we estimate $L_X/L_Y\simeq\mathcal R_{X/Y}^{\mathcal B}$. 
For the full-spectrum comparison, $\mathcal B=[0,1]$, the EH spectra give $\mathcal R_{\rm AM/CR}^{[0,1]}\simeq1.053$, while $\cos(\mathrm{AM},\mathrm{CR})\simeq0.992$. 

\begin{figure}[tb]
    \centering
    \includegraphics[width=0.45\textwidth]{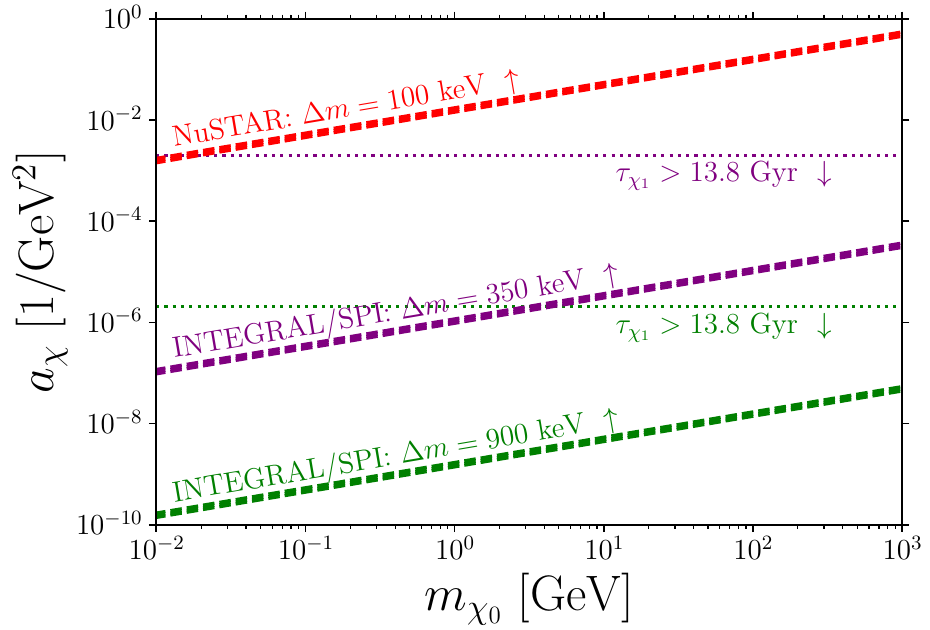}
    
    \vspace{10pt}
    
    \includegraphics[width=0.45\textwidth]{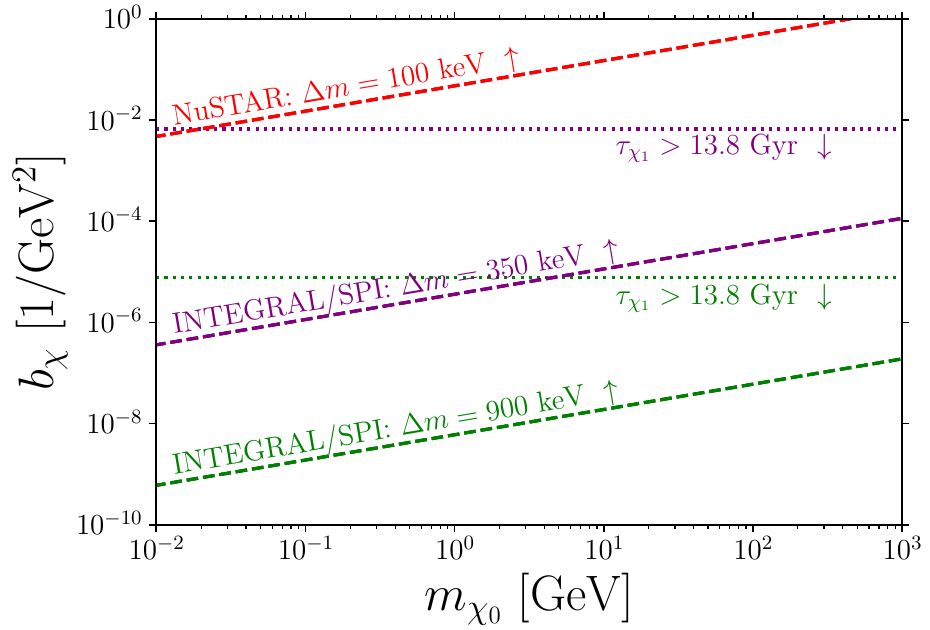}
    \caption{
            X-ray limits on AM (top) and CR (bottom) iDM from INTEGRAL/SPI and NuSTAR for the indicated mass splittings, with one nonzero transition coefficient at a time.
            The $\Delta=900\,\mathrm{keV}$ benchmark is also used in \cref{fig:main_results}.
            The limits strengthen rapidly with increasing splitting because $\Gamma_{3\gamma}\propto\Delta^{13}$ in the EH limit.
            The dotted lines mark a lifetime equal to the age of the Universe.
            Above these lines, decay depletion reduces the surviving excited fraction and weakens the X-ray constraints.
        }
    \label{fig:dim6_INTEGRAL_NuSTAR}
\end{figure}

Both AM and CR spectra integrate to three photons per decay and have $\langle x\rangle \simeq 1/3$, while finite-electron-mass corrections distort the normalized AM and CR spectra in a similar manner.
We therefore adopt
\begin{equation}
    L_{\rm AM}(\Delta) \simeq \mathcal R_{\rm AM/CR}\,L_{\rm CR}(\Delta)
    \,,
    \quad
    \mathcal R_{\rm AM/CR} \simeq 1.053
    \,,
    \label{eq:AM_lifetime_recast}
\end{equation}
which gives 
\be
    L_{\rm AM}(900\,\mathrm{keV})\simeq 7.8\times10^{23}\,\mathrm{GeV}\,\mathrm{s}
\,.
\ee
We translate these lifetime limits into the Wilson-coefficient bounds shown in \cref{fig:dim6_INTEGRAL_NuSTAR}.
For the AM recast, we assign a $20\%$ uncertainty to the lifetime--mass limit, corresponding to approximately $10\%$ in the coefficient because the bound scales as $L^{-1/2}$.
We define the full-rate correction factors
\begin{equation}
    K_X(\Delta)=\frac{\Gamma_X}{\Gamma_X^0}
    \,,
    \label{eq:KX_definition}
\end{equation}
and $L_{X,24}= L_X/(10^{24}\,\mathrm{GeV}\,\mathrm{s})$.  Combining \cref{eq:Gamma_num,eq:CR_lifetime_bound,eq:AM_lifetime_recast} yields
\begin{align}
    |a_\chi| &\lesssim3.09\times10^{-9}\,\mathrm{GeV}^{-2}
    \left[ \frac{m_{\chi_0}/\mathrm{GeV}} {2f_1K_{\rm AM}(\Delta)L_{{\rm AM},24}(\Delta)} \right]^{1/2}
    \nonumber\\
    & \quad \left(\frac{900\keV}{\Delta}\right)^{13/2}
    \,,
    \label{eq:achi_xray_bound}
    \\
    |b_\chi| &<1.02\times10^{-8}\,\mathrm{GeV}^{-2}
    \left[ \frac{m_{\chi_0}/\mathrm{GeV}} {2f_1K_{\rm CR}(\Delta)L_{{\rm CR},24}(\Delta)} \right]^{1/2}
    \nonumber\\
    & \quad \left(\frac{900\keV}{\Delta}\right)^{13/2}
    \label{eq:bchi_xray_bound}
    \,.
\end{align}
At $\Delta=900\,\mathrm{keV}$, where $K_{\rm AM}\simeq 5.3$ and $K_{\rm CR}\simeq 3.9$, these become, for $f_1=1/2$,
\be
    |a_\chi| &\lesssim 1.5 \times10^{-9}\,\mathrm{GeV}^{-2} \sqrt{\frac{m_{\chi_0}}{\mathrm{GeV}}}
    \,,
    \\
    |b_\chi| &\lesssim 6.0\times10^{-9}\,\mathrm{GeV}^{-2} \sqrt{\frac{m_{\chi_0}}{\mathrm{GeV}}}
    \,.
    \label{eq:operator_limits_900}
\ee
The larger full-rate correction in the AM case strengthens the bound on $a_\chi$, but it does not enter the spectral rescaling in \cref{eq:AM_lifetime_recast} -- at fixed physical lifetime, only the normalized photon spectrum determines the signal template.

While our main interest is the benchmark with $\Delta=900$ keV, where INTEGRAL/SPI limit is dominant, for lower mass splittings NuSTAR stray-light constraint of Ref.~\cite{Jeesun:2026ryo}, which was obtained from the $3$--$18\,\keV$ spectrum using an ${\rm SNR}=3$ Asimov criterion, is stronger.
At $\Delta=100\,\keV$, Fig.~9 of that reference gives (their statistical criterion differs from the INTEGRAL/SPI credible limit quoted above)
\be
    L_{3\gamma}^{\rm NuSTAR}&\simeq1.16\times10^{23}\,\GeV\,{\rm s}\,,
    \\
    L_{2\gamma}^{\rm NuSTAR}&\simeq1.30\times10^{22}\,\GeV\,{\rm s}\,,
\ee
for the vector-mediated three-photon and scalar-mediated two-photon decays, respectively, assuming $f_1=1/2$. These limits apply directly to the CR and scalar-Rayleigh spectra. 
For the AM and pseudoscalar spectra, we obtain limits by recasting in a similar way to the INTEGRAL/SPI -- we use \cref{eq:flat_shape_cosine,eq:normaliza} with $\mathcal B=[0.03,0.18]$.
Using the EH spectra, at $\Delta=100\,\mathrm{keV}$, this prescription gives  $\mathcal R_{\rm AM/CR}^{\rm band}\simeq0.821$ and $\mathcal R_{P/S}^{\rm band}\simeq1.834$. 
The sizable pseudoscalar rescaling is a bandpass effect: NuSTAR observes only $0.03<x<0.18$, where the pseudoscalar spectrum is enhanced relative to the scalar spectrum.

Using $K_{\rm AM}\simeq1.014$ and $K_{\rm CR}\simeq1.012$, the resulting NuSTAR bounds are
\be
    |a_\chi| &\lesssim 1.6\times10^{-2}\,\GeV^{-2}
    \sqrt{\frac{m_{\chi_0}}{\GeV}}\,, \\
    |b_\chi| &\lesssim 4.8\times10^{-2}\,\GeV^{-2}
    \sqrt{\frac{m_{\chi_0}}{\GeV}}\,,
    \\
    |C_S|,|\widetilde C_S| &\lesssim 2.9\times10^{-8}\,\GeV^{-3}
    \sqrt{\frac{m_{\chi_0}}{\GeV}}\,,
    \\
    |C_P|,|\widetilde C_P| &\lesssim 7.3\times10^{-4}\,\GeV^{-3}
    \left(\frac{m_{\chi_0}}{\GeV}\right)^{3/2}
    \,.
\ee
These limits are shown in \cref{fig:dim6_INTEGRAL_NuSTAR} for dimension-6 operators and in \cref{fig:Rayleigh_INTEGRAL_NuSTAR} for dimension-7 operators discussed in the Appendix~\ref{app:Ray}.
At fixed DM mass, the AM and CR coefficient bounds at $100\,\mathrm{keV}$ are approximately seven orders of magnitude weaker than the INTEGRAL/SPI bounds at $900\,\mathrm{keV}$.
This difference is driven primarily by the $\Delta^{13}$ dependence of the width, with additional dependence on the instrumental bandpass and statistical sensitivity.
Such large coefficients lie far above the freeze-in targets near their minima and require particular care with terrestrial constraints and perturbative ultraviolet matching.
The $100\,\mathrm{keV}$ comparison is therefore kept separate from the fixed-$900\,\mathrm{keV}$ results shown in \cref{fig:main_results}.

\subsection{Searches at SND@LHC}
\label{sec:SND}
Proton collisions produce an energetic iDM flux that can be searched for through scattering at IF detectors~\cite{Jodlowski:2023ohn,Dienes:2023uve}.
In this work, we focus on prospects of SND@LHC through scattering in its tungsten target; searches at other IF experiments have been studied in~\cite{Jodlowski:2023ohn,Kling:2022ykt,Chu:2020ysb}.
Located in the TI18 tunnel approximately $480~\mathrm{m}$ downstream of the ATLAS interaction point, SND@LHC covers the off-axis region $7.2\lesssim\eta\lesssim8.4$~\cite{SHiP:2020sos,Ahdida:2750060}. 
Its target and vertex detector comprise five approximately $40\times40~\mathrm{cm}^2$ emulsion-cloud-chamber modules, formed from alternating emulsion films and $1~\mathrm{mm}$ tungsten plates and interleaved with scintillating-fiber planes. Together with the upstream veto and downstream calorimeter and muon system, this configuration permits searches for isolated electron recoils and neutral current nuclear recoils.

For the couplings of interest and $\Delta<2m_e$, both states are effectively stable on collider scales and reach the detector.
Their scattering signatures resemble those of elastic EM form factor DM~\cite{Chu:2020ysb,Kling:2022ykt}.
SND@LHC can probe endothermic upscattering, $\chi_0T\to\chi_1T$, and exothermic downscattering, $\chi_1T\to\chi_0T$, with $T=e,\mathcal N$. 
Following Refs.~\cite{Biswas:2026ahs,Kling:2022ykt}, we require the incident trajectory to intersect the finite target area, impose $7.2<\eta_\chi<8.4$ together with the corresponding azimuthal acceptance, and require a visible recoil energy $E_R>100~\mathrm{MeV}$.
The electron channel additionally requires an isolated electromagnetic recoil without an incoming veto signal, reconstructed muon, or accompanying hadronic activity. 
Since $\Delta<2m_e$ is negligible relative to the typical incident and recoil energies, these elastic-search cuts remain applicable. 
We retain exact inelastic kinematics in the differential cross sections and integration limits, using Ref.~\cite{Jodlowski:2023ohn}, to treat upscattering thresholds and recoil endpoints consistently.

We include two complementary production mechanisms. At small masses, the accepted flux is dominated by two-body decays of neutral vector mesons,
\begin{equation}
    V\to\gamma^*\to\chi_0\chi_1\,, \,\,
    V=\rho,\,\omega,\,\phi,J/\psi,\,\psi(2S),\,\Upsilon(nS)\,,
\end{equation}
while the corresponding three-body decays of pseudoscalar mesons are subleading. We obtain the energy and angular distributions of the parent mesons from the pre-generated spectra supplied with \texttt{FORESEE}~\cite{Kling:2021fwx}. The light-meson spectra are generated with EPOS-LHC~\cite{Pierog:2013ria} through the CRMC~\cite{CRMC} interface, while the charmonium and bottomonium spectra are generated with \texttt{Pythia~8} and tuned to forward LHCb measurements~\cite{Pierog:2013ria,Sjostrand:2006za,Sjostrand:2014zea}. 
We then use \texttt{FORESEE} to simulate the decay kinematics, boost the dark states to the laboratory frame, and impose the SND@LHC geometric acceptance and scattering cuts. We use formulas for vector-meson branching ratios from Ref.~\cite{Jodlowski:2023ohn}, and we retain the exact dependence on iDM-state masses.

At larger DM masses, vector-meson decays become phase-space suppressed and eventually close. We then include Drell--Yan production, $q\bar q \to\chi_0\chi_1$.
Using our implementation of the hypercharge-completed AM and CR iDM interactions in \texttt{FeynRules}~\cite{Alloul:2013bka}, we generated $pp\to\chi_0\chi_1$ spectrum at $\sqrt{s}=14~\mathrm{TeV}$ via \texttt{MadGraph5\_aMC@NLO}~\cite{Alwall:2014hca}. 
We pass the resulting spectrum to \texttt{FORESEE} to apply the geometric acceptance and evaluate the energy-dependent scattering probability.

Meson decays dominate at small masses, while Drell--Yan production controls the reach after the relevant meson channels become suppressed or kinematically closed.
Since Drell--Yan production can probe virtualities well above the meson masses, the EFT interpretation requires the invariant mass of the produced pair to remain below the heavy DP mass.
We restrict the EFT interpretation of Drell--Yan production to parameter points satisfying this scale separation.

For each accepted state, we compute the scattering probability.
The expected number of events is
\be
\label{eq:SND_events}
    N_{\mathrm{sig}} &= \sum_{i=0,1}\sum_{T=e,\mathcal N} \int dE_{\chi_i}\, \frac{dN_{\chi_i}^{\mathrm{acc}}}{dE_{\chi_i}}\, n_T \, L_{\mathrm{det}} 
    \times\nonumber\\
    & \times\int_{E_R^{\mathrm{min}}}^{E_R^{\mathrm{max}}}dE_R\, \frac{d\sigma_{\chi_iT\to\chi_jT}}{dE_R}
    \,,
    \qquad j\neq i
    \,,
\ee
where $dN_{\chi_i}^{\mathrm{acc}}/dE_{\chi_i}$ is the accepted flux from both production modes, $n_T$ is the target number density, $E_R^{\mathrm{min}}$ is the energy cut following~\cite{Biswas:2026ahs}, $E_R^{\mathrm{max}}$ is the kinematically-allowed maximal energy, and $L_{\mathrm{det}}$ is the effective tungsten length.
For electrons, $n_e=Z_{\mathrm W}n_{\mathrm W}$ with $Z_{\mathrm W}=74$. For nuclear scattering, we include atomic screening and the electric nuclear form factor, which regulate the small-momentum region and the loss of coherence at large momentum transfer~\cite{Schiff:1953yzz,Tsai:1973py,Helm:1956zz}. 
Since both production and scattering are quadratic in the relevant Wilson coefficient, the event yield scales approximately as $N_{\mathrm{sig}}\propto a_\chi^4$ or $b_\chi^4$ away from thresholds.

The momentum dependence of the AM and CR vertices cancels the virtual-photon propagator. 
Their recoil spectra therefore lack the strong small-$q^2$ enhancement of millicharge and dipole interactions~\cite{Jodlowski:2023ohn,Chu:2020ysb}.
The sensitivity depends more strongly on the accepted flux and the upper recoil endpoint, which limits the benefit of lowering an already small recoil threshold.

The dominant irreducible background to the electron channel is neutrino--electron scattering. Neutrino interactions on nuclei can also mimic either channel when accompanying particles are not reconstructed, while neutral-hadron backgrounds require a detector-level treatment. 
Following Ref.~\cite{Biswas:2026ahs}, we assume that the stated topological and visibility requirements reduce the residual background to a negligible level and take unit efficiency above threshold. We do not impose a timing cut because the accepted dark states are highly relativistic. 
For zero observed events, the projected $90\%$ C.L. exclusion is defined by $N_{\mathrm{sig}}=2.3$. 
We use an integrated luminosity of $3~\mathrm{ab}^{-1}$ for the high-luminosity LHC and compare the SND@LHC projection with the FLArE reach of Ref.~\cite{Kling:2022ykt}.
SND@LHC is less sensitive than FLArE, in part because of its off-axis location.
Both projections lie at larger couplings than the collider bounds considered here, reflecting the contact nature of the interaction and the quartic dependence of the signal yield on the Wilson coefficient.

\subsection{Direct detection searches}
\label{sec:DD}
The cosmological history of our model yields sizable present-day abundances of both $\chi_0$ and $\chi_1$, allowing DD experiments to constrain the model through the exothermic downscattering process $\chi_1 T\to\chi_0 T$. 
The inverse endothermic process is kinematically suppressed or forbidden for the splittings and halo velocities relevant here.
The high-velocity population induced by the Large Magellanic Cloud can extend sensitivity to endothermic scattering~\cite{Reynoso-Cordova:2024xqz}, but does not remove this suppression for our main benchmark.

Following the inelastic nonrelativistic EFT construction of Ref.~\cite{Barello:2014uda}, we consider $\chi_i(\mathbf p_i)N(\mathbf k_i)\to \chi_f(\mathbf p_f)N(\mathbf k_f)$ and use the momentum-transfer convention
\begin{equation}
  \mathbf q = \mathbf p_f-\mathbf p_i = \mathbf k_i-\mathbf k_f\,, \qquad \delta_{if}=m_f-m_i\,.
\end{equation}
Thus, $\delta_{01}=\Delta$ for endothermic up-scattering and $\delta_{10}=-\Delta$ for exothermic down-scattering.  
The Galilean-invariant transverse velocity is
\begin{equation}
  \mathbf v^{\perp}_{\mathrm{inel}} = \mathbf v + \frac{\mathbf q}{2\mu_{iN}} + \frac{\delta_{if}}{\mathbf q^2}\,\mathbf q\,, \quad \mathbf v^{\perp}_{\mathrm{inel}}\!\cdot\mathbf q=0\,, \quad \mu_{iN}=\frac{m_i m_N}{m_i+m_N}
  \,.
\end{equation}
Here $\mathbf v$ is the incoming DM--nucleon relative velocity. 
The transition AM and CR interactions involve
\begin{align}
  \mathcal O_1^N &= \mathbb 1\,,
  &
  \mathcal O_8^N &= \mathbf S_\chi\!\cdot\mathbf v^{\perp}_{\mathrm{inel}}\,,
  \nonumber\\
  \mathcal O_9^N &= i\mathbf S_\chi\!\cdot \left(\mathbf S_N\times\frac{\mathbf q}{m_N}\right)\,,
  &
  \mathcal O_{11}^N &= i\mathbf S_\chi\!\cdot\frac{\mathbf q}{m_N}\,.
\end{align}
In the normalization $\mathcal M_{iN\to fN}^{\mathrm{NR}} =\sum_j c_j^{N,if}\mathcal O_j^N$, the nonzero one-nucleon coefficients for the $0\to1$ transition at leading nonrelativistic order are
\begin{align}
  c_1^{N,01} &= i e Q_N b_\chi\,,
  &
  c_8^{N,01} &= 2eQ_N a_\chi\,,
  \nonumber\\
  c_9^{N,01} &= -e g_N a_\chi\,,
  &
  c_{11}^{N,01}(\mathbf q) &= 2i eQ_N a_\chi \frac{m_N\delta_{01}}{\mathbf q^2}
\,.
  \label{eq:nreft-matching}
\end{align}
Here $Q_p=1$, $Q_n=0$, and $g_N$ denotes the full nucleon magnetic $g$ factor, with $g_p\simeq 5.586$ and $g_n\simeq-3.826$.  
The AM charge coupling produces $\mathcal O_8$ and the splitting-dependent $\mathcal O_{11}$ term, while the CR interaction excites the coherent nuclear $M$ response through $\mathcal O_1$; both contribute to the $M$ response.
The nucleon magnetic coupling produces $\mathcal O_9$, including interference with $\mathcal O_8$.

The factors of $i$ in \eqref{eq:nreft-matching} follow from the hermiticity convention for off-diagonal  operators. 
Rates depend on the products
\begin{equation}
  c_j^{N,if}\,c_k^{N',if*}\,,
  \qquad
  c_j^{N,10}=c_j^{N,01*}\,.
\end{equation}
Therefore, $c_1^{N,10}=-i eQ_Nb_\chi$, while $c_{11}^{N,10}$ is obtained from~\eqref{eq:nreft-matching} by replacing $\delta_{01}$ with $\delta_{10}=-\Delta$. 

We evaluate the nuclear recoil spectra with \texttt{WimPyDD}~\cite{Jeong:2021bpl,Kang:2022bxr}.  
For a target nuclide $T$, the minimum incoming speed is
\begin{equation}
  v_{\min}^{i\to f}(E_R) = \frac{1}{\sqrt{2m_T E_R}} \left| \frac{m_T E_R}{\mu_{iT}}+\delta_{if} \right|\,,
  \label{eq:vmin-inelastic}
\end{equation}
where $q=|\mathbf q|=\sqrt{2m_T E_R}$. 
When both dark-sector states are present locally, the upscattering and downscattering contributions are evaluated separately.
For a target mass fraction $\xi_T$, the recoil rate per detector mass is 
\begin{equation}
  \frac{dR_T}{dE_R} = \frac{\xi_T}{m_T} \sum_{(i,f)=(0,1),(1,0)} \frac{\rho_i}{m_i} \int_{v\geq v_{\min}^{i\to f}} d^3v\,f_i(\mathbf v)\,v\, \frac{d\sigma_{iT\to fT}}{dE_R}
  \,.
  \label{eq:two-state-rate}
\end{equation}
Here $f_i(\mathbf v)$ is normalized to unity and $\rho_i=f_i^{\mathrm{loc}}\rho_{\mathrm{DM}}$, where $f_i^{\mathrm{loc}}$ is the local mass fraction of state $i$. 
We obtain the predicted counts by folding each channel with the detector response and acceptance implemented in the corresponding \texttt{WimPyDD} dataset.
We assume that the local state fractions trace their cosmological values.

\begin{figure*}[tb]
    \centering
    \includegraphics[width=0.45\textwidth]{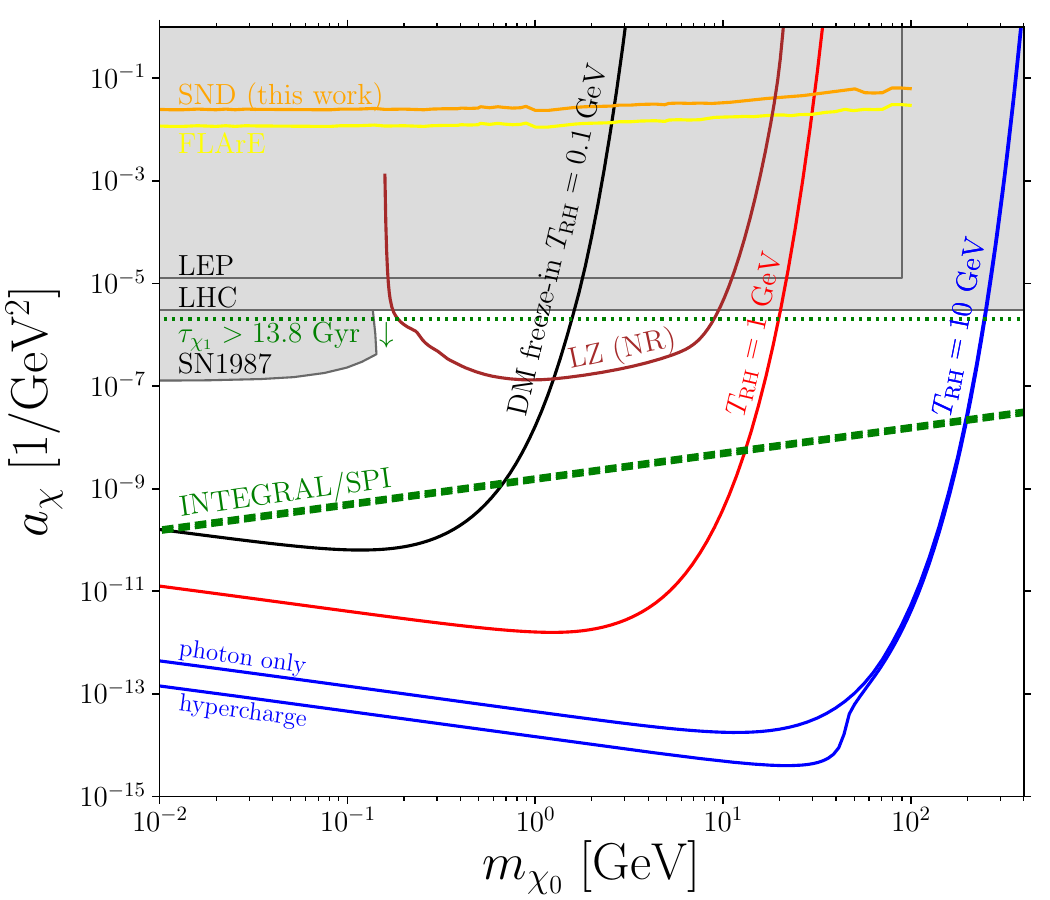}
    \hspace{0.03\textwidth}
    \includegraphics[width=0.45\textwidth]{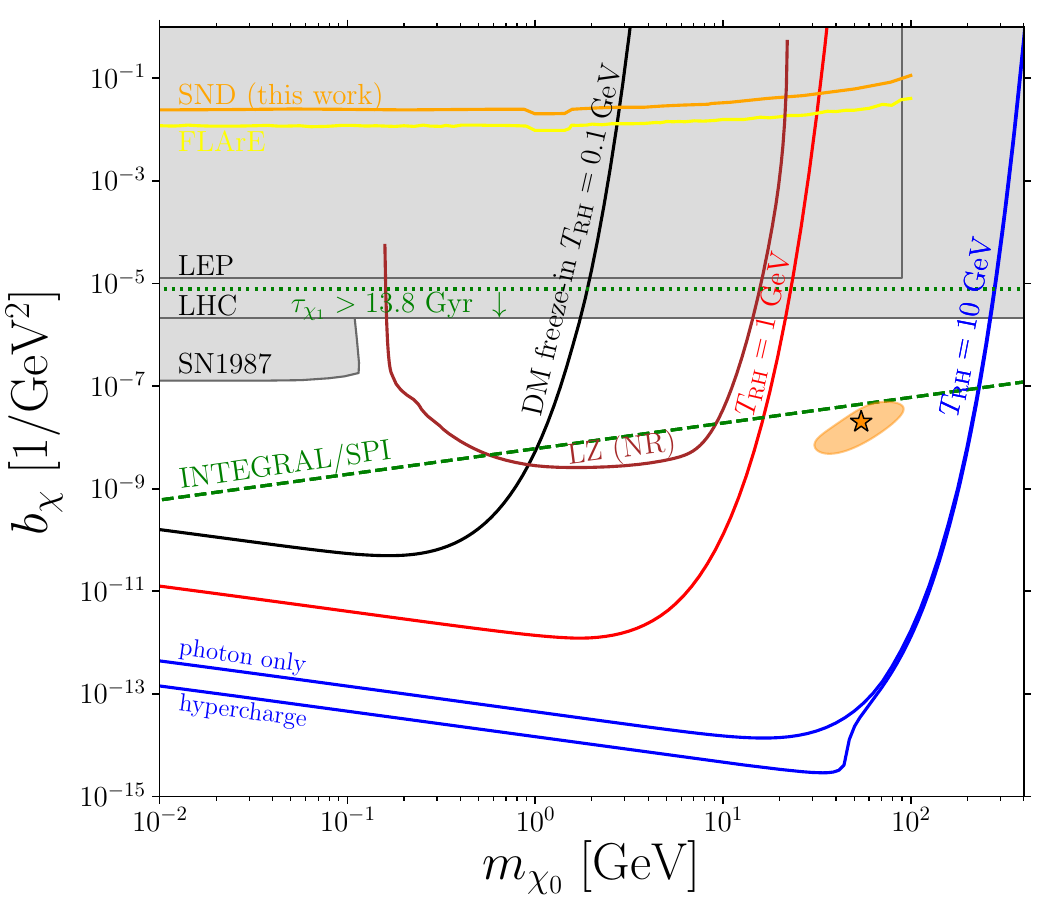}
    \caption{
            Freeze-in targets and experimental constraints for AM iDM (left) and CR iDM (right), with $\Delta=900\,\mathrm{keV}$.
            Black, red, and blue curves give $\Omega_\chi h^2=0.12$ for $T_{\rm RH}=0.1$, $1$, and $10\,\mathrm{GeV}$, respectively.
            Curves labeled ``hypercharge'' include the correlated photon and $Z$ interactions.
            The ``photon only'' curves are reference calculations for the analytical limits.
            Thick green dashed curves show the INTEGRAL/SPI bounds in the long-lived regime, with couplings above excluded.
            The AM recast carries $\sim 10\%$ uncertainty in the coefficient.
            Thin green dotted lines mark $\tau_{\chi_1}=13.8\,\mathrm{Gyr}$.
            Gray regions denote the indicated existing constraints, and brown denotes the LZ nuclear-recoil bound for a present local excited-state fraction $f_1=1/2$.
            Orange and yellow curves show SND@LHC and FLArE projections, respectively.
            In the CR panel, the orange region and star indicate the approximate LZ signal-only likelihood region and representative maximum discussed in \cref{sec:DD}.
        }
    \label{fig:main_results}
\end{figure*}

We use the \texttt{WimPyDD} implementations of LZ~\cite{LZ:2022lsv}, PICO-60~\cite{PICO:2019vsc}, and XENON1T~\cite{XENON:2017vdw}.
As a normalization check, the limit $\delta_{if}\to0$ reproduces the elastic CR and AM rates of Ref.~\cite{Ibarra:2024mpq}.
This check does not constrain the splitting-dependent $\mathcal O_{11}$ contribution because $c_{11}\propto\delta_{if}$. 
Among these nuclear-recoil searches, LZ gives the strongest bound, and is shown in brown in \cref{fig:main_results}.
It corresponds to the tabulated nuclear-recoil acceptance and the effective signal-count limit $N_{\rm sig}\lesssim3.37$.
For $\Delta=900\,\mathrm{keV}$, the strongest sensitivity occurs at masses of order a GeV, where the exothermic recoils fall within the accepted energy range, and weakens at larger masses as the recoil spectrum moves above it. 
The shown brown lines assume the local excited-state fraction to be $f_1=1/2$.
Thus, the actual limits hold only for points satisfying this constraint.
We also note that our recast is simplified compared to  the collaboration's full likelihood, and is separate from the high-energy recoil-candidate analysis below.

\textit{The LZ recoil candidate.}
The CR interaction offers a possible interpretation of the high-energy nuclear-recoil candidate reported by LZ~\cite{LZ:2026axp} via exothermic scattering, $\chi_1 N\to\chi_0 N$; related vector-mediated explanations have been explored in~\cite{deLima:2026shq,Baer:2026fpy}.
For $\Delta=m_{\chi_1}-m_{\chi_0}>0$, the minimum incident speed vanishes at $E_0=\Delta\,m_{\chi_0}/(m_{\chi_0}+m_N)$, allowing a high-energy recoil without requiring the fastest halo particles.

To estimate the relevant mass and coupling scales, we fix $\Delta=900~\mathrm{keV}$ and $f_1=1/2$. We use natural xenon with Helm charge form factors, an exposure of $2.84~\mathrm{tonne\,yr}$, and the Standard Halo Model with $\rho_{\rm DM}=0.3~\mathrm{GeV\,cm^{-3}}$, adopting the halo velocities and approximate main-window acceptance of Ref.~\cite{Baer:2026fpy}.\footnote{We take $(v_0,v_{\rm esc},v_E)=(238,544,254)~\mathrm{km\,s^{-1}}$. The acceptance rises linearly from $0.5$ at $5.4~\mathrm{keV}$ to $0.96$ at $14~\mathrm{keV}$, is constant up to $250~\mathrm{keV}$, and falls linearly to $0.5$ at $270~\mathrm{keV}$. We apply this acceptance to the reconstructed energy after Gaussian smearing with $\sigma_E=\sqrt{23^2+23^2}~\mathrm{keV}$, combining the candidate's quoted statistical and systematic errors in quadrature; we set the sideband acceptance to unity.}
Writing $m=m_{\chi_0}$ and $s(E,m,b_\chi)$ for the expected signal counts per reconstructed recoil energy after smearing and acceptance, we use the simplified likelihood
\begin{equation}
  \begin{aligned}
  \mathcal L_{\rm app}(m,b_\chi) &\propto e^{-\nu(m,b_\chi)}s(E_\star,m,b_\chi)
\,,
  \\
  \nu(m,b_\chi)&=\int_W dE\,s(E,m,b_\chi)
\,,
  \end{aligned}
\label{eq:lzcr-likelihood}
\end{equation}
where $E_\star=248~\mathrm{keV}$ and $W=[5.4,270]\cup[350,650]~\mathrm{keV}$ includes the main window and the empty sideband adopted in Ref.~\cite{deLima:2026shq}. This signal-only prescription assigns the candidate to signal and assumes no additional signal candidates in these windows. 
It gives a representative maximum near $m\simeq55~\mathrm{GeV}$ and $|b_\chi|\simeq2\times10^{-8}~\mathrm{GeV}^{-2}$. Using tabulated nuclear charge responses instead shifts the coupling to approximately $3\times10^{-8}~\mathrm{GeV}^{-2}$, with little change in the preferred mass. This difference reflects the nuclear-response choice, rather than a statistical interval. For other fixed $f_1$, the couplings rescale by $(2f_1)^{-1/2}$.

The shaded region denotes $q=-2\ln(\mathcal L_{\rm app}/\mathcal L_{\rm app,max})<2.3$ and spans approximately $31$--$92~\mathrm{GeV}$ in the Helm calculation. We use this as an illustrative likelihood contour without assigning a calibrated confidence level. At $\Delta=900~\mathrm{keV}$, the INTEGRAL/SPI bound~\cite{Krnjaic:2025zjl}, $|b_\chi|\lesssim6.0\times10^{-9}\sqrt{(m/\mathrm{GeV})/(2f_1)}~\mathrm{GeV}^{-2}$, is compatible with the representative CR normalization near $55~\mathrm{GeV}$. However, the AM interpretation at this mass and splitting requires $|a_\chi|\sim10^{-6}~\mathrm{GeV}^{-2}$, exceeding the corresponding X-ray bound of order $10^{-8}~\mathrm{GeV}^{-2}$ for $f_1=1/2$.

\section{Probing freeze-in iDM}
\label{sec:results}
Having discussed details of freeze-in of iDM and its signatures at X-ray, DD, collider, and IF searches in proceeding two sections, we present the main result of our study -- the prospects of probing the freeze-in region with INTEGRAL/SPI.
We present our combined results in~\cref{fig:main_results}.
The relic-density contours are obtained by solving \cref{eq:fi-coupled-boltzmann} with vanishing initial dark-sector abundance and imposing $\Omega_\chi h^2=0.12$ through \cref{eq:fi-relic-density}~\cite{Planck:2018vyg}.
The excited-state fraction entering the decay flux is taken from the numerical evolution.
It remains close to $f_1=1/2$ because production creates one $\chi_0$ and one $\chi_1$, while conversion is inefficient.
For $\Delta\ll m_{\chi_0}$ and the reheating benchmarks considered here, the freeze-in contours depend only weakly on the splitting and resemble those of the elastic theory.
In contrast, the X-ray limits depend strongly on $\Delta$, as illustrated in \cref{fig:dim6_INTEGRAL_NuSTAR,fig:Rayleigh_INTEGRAL_NuSTAR}.
We therefore keep the splitting fixed throughout the comparison in \cref{fig:main_results}.

At fixed $\Delta$, the relation between a lifetime limit and a Wilson-coefficient limit follows directly from \cref{eq:achi_xray_bound,eq:bchi_xray_bound}. 
Its parametric form is
\begin{equation}
    c_\chi^{\mathrm{X}}
    \propto
    \left[
        \frac{m_{\chi_0}}
        {f_1 K_X(\Delta)L_X(\Delta)}
    \right]^{1/2}
    \Delta^{-13/2}
\,,
\end{equation}
where $c_\chi=a_\chi$ or $b_\chi$, $K_X$ is the full one-loop rate correction defined in \cref{eq:KX_definition}, and $L_X$ is the corresponding limit on $\tau_{\chi_1}m_{\chi_0}$.
At fixed splitting and excited-state fraction, the coefficient bound therefore weakens only as $m_{\chi_0}^{1/2}$.
Its much steeper splitting dependence follows from the $\Delta^{13}$ scaling of the EH three-photon width, modified by the loop correction and the spectral dependence of the observational limit.

For $\Delta=900\,\mathrm{keV}$ and $f_1=1/2$, including the full electron-loop correction gives the following scaling of the bound:
\be
    |a_\chi| &\lesssim1.5\times10^{-9}\,\mathrm{GeV}^{-2} \sqrt{\frac{m_{\chi_0}}{\mathrm{GeV}}}
    \,,
    \\
    |b_\chi| &\lesssim6.0\times10^{-9}\,\mathrm{GeV}^{-2} \sqrt{\frac{m_{\chi_0}}{\mathrm{GeV}}}
\,.
\ee
The stronger AM limit has two origins. 
At equal Wilson coefficients, the AM width exceeds the CR width already in the EH limit, as shown in \cref{eq:Gamma_an}.
The full loop amplitude further enhances the widths by $K_{\rm AM}\simeq5.3$ and $K_{\rm CR}\simeq3.9$ at this splitting.
The CR spectrum coincides with the vector-mediator template, allowing a direct recast.
For AM, the similar normalized spectrum motivates \cref{eq:AM_lifetime_recast}, with an estimated $10\%$ uncertainty in the coefficient bound.
The asymptotic freeze-in scalings explain the high-mass intersections in \cref{fig:main_results}.

At fixed $\Delta$, neglecting variations in the relativistic degrees of freedom, open production channels, and $f_1$, the photon-mediated limits in \cref{eq:fi-rel-scaling,eq:fi-nr-scaling} give
\begin{align}
    \frac{a_\chi^{\mathrm{FI}}}{a_\chi^{\mathrm{X}}}
    &\propto
    \begin{cases}
        m_{\chi_0}^{-1}T_{\mathrm{RH}}^{-3/2}\,,
        &m_{\chi_0}\ll T_{\mathrm{RH}}\,,\\[2mm]
        m_{\chi_0}^{-5/2}
        \exp(m_{\chi_0}/T_{\mathrm{RH}})\,,
        &m_{\chi_0}\gg T_{\mathrm{RH}}\,,
    \end{cases}
    \nonumber\\
    \frac{b_\chi^{\mathrm{FI}}}{b_\chi^{\mathrm{X}}}
    &\propto
    \begin{cases}
        m_{\chi_0}^{-1}T_{\mathrm{RH}}^{-3/2}\,,
        &m_{\chi_0}\ll T_{\mathrm{RH}}\,,\\[2mm]
        T_{\mathrm{RH}}^{1/2}m_{\chi_0}^{-3}
        \exp(m_{\chi_0}/T_{\mathrm{RH}})\,,
        &m_{\chi_0}\gg T_{\mathrm{RH}}\,.
    \end{cases}
\end{align}
For the benchmarks and mass range shown in \cref{fig:main_results}, INTEGRAL/SPI intersects only the Boltzmann-suppressed, high-mass branch of each freeze-in target.
For larger reheating temperatures, the relativistic target moves to smaller couplings and its low-mass intersection is displaced toward masses comparable to or below the chosen splitting, outside the controlled near-degenerate regime.
On this branch, the coupling required to compensate for suppressed pair production grows exponentially with $m_{\chi_0}/T_{\mathrm{RH}}$, eventually exceeding the decay limit.
The AM target is more strongly constrained because its three-photon width is larger and its production is $p$-wave suppressed.
In the photon-mediated nonrelativistic limit, the latter gives $a_\chi^{\mathrm{FI}}/b_\chi^{\mathrm{FI}}\simeq\sqrt{m_{\chi_0}/T_{\mathrm{RH}}}$.

Although the relativistic scaling predicts an increasing ratio of target coupling to X-ray limit toward smaller masses, no low-mass intersection occurs in the displayed parameter range.
Extrapolating this behavior toward $m_{\chi_0}\sim\Delta$ would leave the near-degenerate regime used in the calculation.
The X-ray constraint thus truncates the high-mass branch at each reheating temperature, while the displayed relativistic branch remains below the limit.
The correlated $Z$ interaction modifies the numerical targets, particularly for $T_{\mathrm{RH}}=10\,\mathrm{GeV}$, without changing this conclusion.

At this fixed splitting, LZ provides the strongest terrestrial constraint over approximately $0.2$--$10\,\mathrm{GeV}$ in both models.
In the CR case, its exothermic nuclear-recoil bound also improves on INTEGRAL/SPI over part of the sub-GeV to several-GeV range.
For AM, the X-ray limit remains stronger.
The LEP~\cite{L3:2003yon}, LHC~\cite{Gao:2013vfa}, and SN1987A~\cite{Chu:2018qrm} bounds lie at larger couplings than the INTEGRAL/SPI limit over most of the displayed range, as do the projected SND@LHC and FLArE reaches.
These comparisons apply to $\Delta=900\,\mathrm{keV}$.
At smaller splittings, the rapid weakening of the decay limits changes their relative importance, as discussed in \cref{sec:xray_recast}.

\section{Summary}
\label{sec:conclusions}
We have studied freeze-in production and experimental probes of iDM whose leading coupling to the SM is a dimension-6 transition AM or CR operator.
Both interactions arise at tree level when the heavy DP of a broken $U(1)_D$ vector portal is integrated out.
The sign of the dark-fermion mass-matrix determinant selects an axial or vector transition current, while a residual discrete gauge symmetry stabilizes $\chi_0$.
In the dominantly off-diagonal limit, with Higgs-portal effects subdominant, the same dimension-6 interaction controls production, scattering, and decay.

In the heavy-mediator regime, the photon-mediated pair-production cross section grows with center-of-mass energy. Freeze-in is therefore UV dominated, in contrast to the light-vector-mediator regime of Ref.~\cite{Krnjaic:2025zjl}, and the relic-density target depends sensitively on the reheating temperature. Our numerical calculation includes the correlated $Z$ interaction required by the hypercharge completion and reproduces the analytical photon-mediated limits in their respective domains of validity. AM and CR production coincides in the relativistic limit, while the AM target requires a larger coupling near threshold because of $p$-wave suppression.

For $\Delta<2m_e$, the leading visible decay in our benchmarks is $\chi_1\to\chi_0+3\gamma$. We derived its width and photon spectrum in the EH limit and evaluated the full one-loop amplitude numerically with \texttt{Package-X}. Near the dielectron threshold, the full loop dependence is essential. At $\Delta=900\,\mathrm{keV}$, it enhances the AM and CR widths by factors of approximately $5.3$ and $3.9$, respectively. The CR spectrum coincides with the vector-mediator template of Ref.~\cite{Krnjaic:2025zjl}, whereas the similar AM spectrum permits an approximate recast. 
For this splitting and the present excited-state fraction $f_1=1/2$, the 16-year INTEGRAL/SPI analysis implies
$|a_\chi|\lesssim1.5\times10^{-9}\,\mathrm{GeV}^{-2}\sqrt{m_{\chi_0}/\mathrm{GeV}}$ and $|b_\chi|\lesssim6.0\times10^{-9}\,\mathrm{GeV}^{-2}\sqrt{m_{\chi_0}/\mathrm{GeV}}$.
The AM coefficient bound carries an estimated $10\%$ recasting uncertainty. At $\Delta=100\,\mathrm{keV}$, the NuSTAR limits on the AM and CR coefficients are approximately seven orders of magnitude weaker, primarily because of the steep splitting dependence of the three-photon width. We also derived the two-photon widths, spectra, and X-ray limits for transition Rayleigh operators and identified the conditions under which the DH contribution to the decay is subdominant.

The surviving excited-state population also permits exothermic nuclear scattering. LZ provides the leading terrestrial sensitivity over part of the GeV mass range and, for CR iDM, can exceed the X-ray sensitivity. We examined a possible CR interpretation of the reported high-energy recoil candidate using an approximate signal-only likelihood. At $\Delta=900\,\mathrm{keV}$ and $f_1=1/2$, it favors a representative mass near $55\,\mathrm{GeV}$ and a coefficient of order $(2$--$3)\times10^{-8}\,\mathrm{GeV}^{-2}$, compatible with the INTEGRAL/SPI decay bound. 
The AM coefficient required at the same mass and splitting exceeds its decay bound.

For the three reheating benchmarks studied, INTEGRAL/SPI probes the Boltzmann-suppressed high-mass branch of the freeze-in targets at $\Delta=900\,\mathrm{keV}$, while the relativistic branches remain below the decay limit; AM is more strongly constrained because its decay is faster and its near-threshold production is suppressed. 
Therefore, X-ray observations provide a powerful test of the transition interactions and the reheating temperature required to produce the observed relic abundance of iDM.

\begin{acknowledgments}
This work was supported by the National Natural Science Foundation of China (NNSFC) under grants No.~12335005, No.~12575118, and the Special funds for postdoctoral overseas recruitment, Ministry of Education of China. 
\end{acknowledgments}

\appendix

\section{Dark Higgs and Rayleigh operators}
\label{app:Ray}

The main analysis assumes that Higgs-portal contributions are subdominant. In this appendix, we match DH exchange onto dimension-7 transition Rayleigh operators and derive the associated two-photon widths, spectra, and X-ray constraints. 
We also make explicit the parameter dependence of the competition between two- and three-photon decays.

\subsection{Dark Higgs contribution}
The relative importance of the dark Higgs and the dark photon in the model of \cref{sec:UV} depends on their masses, the Higgs-portal coupling, and the transition Yukawa coupling. A dark Higgs much heavier than the dark photon, for example, was considered in Ref.~\cite{Krnjaic:2025zjl}. Here we retain its leading low-energy contribution and determine when it can be neglected in the excited-state decay.

The coupling of the dark radial mode to the Weyl fermions is determined by
\begin{equation}
    Y_\phi=\frac{\partial\mathcal M}{\partial v_D} =\frac{1}{v_D}\begin{pmatrix}f_\eta&0\\0&f_\xi\end{pmatrix}
    \,.
\end{equation}
After diagonalizing the mass matrix and rephasing negative-mass eigenstates, the magnitude of the transition coupling is
\begin{equation}
    |g_{\phi01}|=\frac{|\delta f\sin2\theta|}{2v_D}
    \,.
\end{equation}
The transition bilinear is scalar for $\det\mathcal M>0$ and pseudoscalar for $\det\mathcal M<0$. In the latter case, the rephasing that makes the negative fermion mass positive also turns the real off-diagonal Yukawa coupling into a purely imaginary one. Thus, the transition coupling vanishes when $f_\eta=f_\xi$.

The Higgs-portal coupling $\lambda_{H\Phi}$ leads to mixing with the SM Higgs. Denoting the scalar mixing angle by $\theta_h$, the small-mixing limit gives
\begin{equation}
    \sin\theta_h\simeq\frac{\lambda_{H\Phi}\,v\,v_D}{m_\phi^2-m_h^2}
    \,.
\end{equation}
Here $m_h$ and $m_\phi$ denote the physical scalar masses, and the approximation requires $|\lambda_{H\Phi}vv_D|\ll|m_\phi^2-m_h^2|$. For $m_\phi^2\gg m_h^2$, this reduces to $\sin\theta_h\simeq\lambda_{H\Phi}vv_D/m_\phi^2$.

Of particular importance is the two-photon coupling inherited from the SM Higgs. 
Writing $h$ for the SM Higgs field before scalar mixing, the effective two-photon interaction is
\begin{equation}
    \mathcal L_{h\gamma\gamma} = \frac{\alpha A_\gamma(q^2)}{8\pi v}\, h\,F_{\mu\nu}F^{\mu\nu}
\,,
\end{equation}
where $q^2\lesssim\Delta^2$. 
At leading order in the low-energy matching, including the light-quark and gluon contributions through the hadronic theory, $A_\gamma(0)=50/27$~\cite{Fradette:2018CosmologicalBeamDump}. 
Replacing $A_\gamma(q^2)$ by this value requires $q^2\ll m_e^2$.
Below the dielectron threshold, the relevant momentum dependence is
\begin{equation}
    A_\gamma(q^2)=\frac{50}{27} +A_{1/2}\!\left(\frac{q^2}{4m_e^2}\right)-\frac{4}{3} +\mathcal O\!\left(\frac{q^2}{m_\mu^2},\frac{q^2}{m_\pi^2}\right)
    \,,
\end{equation}
where
\begin{equation}
    A_{1/2}(\tau) =\frac{2}{\tau^2} \left[\tau+(\tau-1)\arcsin^2\!\sqrt{\tau}\right]
    \,,
    \qquad A_{1/2}(0)=\frac{4}{3}
    \,.
\end{equation}

Integrating out both scalar mass eigenstates generates $C_S$ in the AM branch and $\widetilde C_P$ in the CR branch. The inherited two-photon vertex supplies the loop suppression. 
Their propagators enter through the combination
$\sin\theta_h\cos\theta_h \left(\frac{1}{m_h^2}-\frac{1}{m_\phi^2}\right) =\frac{\lambda_{H\Phi}vv_D}{m_h^2m_\phi^2}$.
For $q^2\ll m_h^2,m_\phi^2$, this gives the zero-momentum Wilson coefficient 
\begin{equation}
    C_{\gamma\gamma}(0) \simeq -\frac{\lambda_{H\Phi}\alpha A_\gamma(0) \,\delta f\sin2\theta} {16\pi m_\phi^2m_h^2}
\,.
\end{equation}
Here $C_{\gamma\gamma}=C_S$ in the AM branch and $C_{\gamma\gamma}=\widetilde C_P$ in the CR branch, up to an overall sign set by the fermion-field conventions.
Scalar exchange also induces interactions with other SM fields, proportional to their Higgs couplings. We assume that these contributions to production and scattering are negligible in the main benchmarks. Below we derive the separate condition for the two-photon decay to remain subdominant.

\subsection{Two-photon decay widths}
The Rayleigh operators mediate $\chi_1\to\chi_0+2\gamma$ at tree level in the effective theory. The spin-averaged squared amplitude depends only on the diphoton invariant mass $k^2$, allowing an analytical phase-space integration. At LO in $\Delta/m_{\chi_0}$, we obtain
\begin{align}
    \frac{d\Gamma^0_{C_S/\widetilde C_S}}{dk^2} &\simeq \frac{|C_S/\widetilde C_S|^2}{8\pi^3}\, k^4\sqrt{\Delta^2-k^2}\,,
    \nonumber\\[1mm]
    \frac{d\Gamma^0_{C_P/\widetilde C_P}}{dk^2} &\simeq \frac{|C_P/\widetilde C_P|^2}{32\pi^3 m_{\chi_0}^2}\, k^4(\Delta^2-k^2)^{3/2}
\,,
\end{align}
with $0 < k^2 < \Delta^2$.
Integrating over $k^2$ gives
\begin{align}
    \label{eq:G_Rayleigh}
    \Gamma^0_{C_S/\widetilde C_S} &\simeq \frac{2|C_S/\widetilde C_S|^2}{105\pi^3}\,\Delta^7\,,
    \nonumber\\[1mm]
    \Gamma^0_{C_P/\widetilde C_P} &\simeq \frac{|C_P/\widetilde C_P|^2}{630\pi^3 m_{\chi_0}^2}\,\Delta^9
\,.
\end{align}
For the dark-Higgs-induced interaction, finite-momentum corrections follow from retaining $A_\gamma(q^2)$ in the decay amplitude.
As shown in \cref{fig:beyond_HE}, their contributions are small.

Normalizing to $\Delta=100\,\mathrm{keV}$, the widths are
\be
    \label{eq:G_Rayleigh_num}
    \Gamma^0_{C_S/\widetilde C_S}  &\simeq 9.333\times 10^{-8}\,\mathrm{s}^{-1} \left( \frac{\left|C_S/\widetilde C_S\right|} {1\,\mathrm{GeV}^{-3}} \right)^2 \left( \frac{\Delta}{100\,\mathrm{keV}} \right)^7
    \,,
    \\
    \Gamma^0_{C_P/\widetilde C_P}  &\simeq 7.778\times 10^{-17}\,\mathrm{s}^{-1} \left( \frac{\left|C_P/\widetilde C_P\right|} {1\,\mathrm{GeV}^{-3}} \right)^2 
    \\
    &\,\,\,\, \left( \frac{\Delta}{100\,\mathrm{keV}} \right)^9  \left( \frac{1\,\mathrm{GeV}}{m_{\chi_0}} \right)^2
    \,.
\ee
The milder splitting dependence of the Rayleigh widths reflects the local two-field-strength interaction and the three-body phase space. The AM and CR channels instead involve the EH four-field-strength interaction and four-body phase space.

For the DH-induced coefficients given above, the corresponding $\chi_1\to \chi_0 + 2\gamma$ widths are
\begin{equation}
    \Gamma_{2\gamma}^{\rm AM} \simeq \frac{\lambda_{H\Phi}^2\alpha^2|A_\gamma(0)|^2 (\delta f)^2\sin^22\theta} {13440\pi^5m_\phi^4m_h^4} \Delta^7
\,,
\end{equation}
and
\begin{equation}
    \Gamma_{2\gamma}^{\rm CR} \simeq \frac{\lambda_{H\Phi}^2\alpha^2|A_\gamma(0)|^2 (\delta f)^2\sin^22\theta} {161280\pi^5m_\phi^4m_h^4m_{\chi_0}^2} \Delta^9
\,.
\end{equation}
The additional suppression in the CR branch follows from the nonrelativistic pseudoscalar transition matrix element.
Comparing these expressions with the EH three-photon widths in \cref{eq:Gamma_an}, the three-photon channel dominates for $|\lambda_{H\Phi}|<\lambda_{H\Phi}^{\rm crit}$,
where
\begin{equation}
    \lambda_{H\Phi}^{\rm crit,AM} \simeq 0.043\,\frac{\alpha}{|A_\gamma(0)|} \frac{|a_\chi|m_\phi^2m_h^2\Delta^3} {|\delta f\sin2\theta|m_e^4}
\,,
\end{equation}
and
\begin{equation}
    \lambda_{H\Phi}^{\rm crit,CR} \simeq 0.045\,\frac{\alpha}{|A_\gamma(0)|} \frac{|b_\chi|m_\phi^2m_h^2m_{\chi_0}\Delta^2} {|\delta f\sin2\theta|m_e^4}
\,.
\end{equation}
Defining $\kappa=|\cot2\theta|$ and taking the respective pseudo-Majorana and pseudo-Dirac limits gives
\begin{align}
    \lambda_{H\Phi}^{\rm crit}
    &\simeq
    \begin{cases}
        3.13 & (\mathrm{AM})\\
        1.64 & (\mathrm{CR})
    \end{cases}
    \nonumber\\
    &\quad\times
    \left(\frac{\epsilon g_D}{10^{-8}}\right)
    \left(\frac{m_\phi}{1\,\mathrm{TeV}}\right)^2
    \left(\frac{1\,\mathrm{TeV}}{m_{A'}}\right)^2
    \nonumber\\
    &\quad\times
    \left(\frac{\Delta}{900\,\mathrm{keV}}\right)^2
    \left(\frac{0.1}{\kappa}\right)
    \,,
\end{align}
where we used $m_h=125\,\mathrm{GeV}$, $\alpha^{-1}=137.036$, $A_\gamma(0)=50/27$, and $\sin2\theta\simeq1$. 
The common $\Delta^2$ scaling holds when $\kappa$ is held fixed: $|\delta f|\simeq\kappa\,\Delta$ in the pseudo-Majorana limit, while $|\delta f|\simeq2\kappa\,m_{\chi_0}$ in the pseudo-Dirac limit.

These estimates use the EH three-photon width and the zero-momentum Higgs form factor. Their numerical values near the dielectron threshold are therefore indicative. Including finite-momentum corrections to both widths changes the critical coupling by the square root of the ratio of their correction factors. In the main analysis, we assume $|\lambda_{H\Phi}|\ll\lambda_{H\Phi}^{\mathrm{crit}}$, ensuring that two-photon decay is subdominant, together with negligible scalar-mediated production and scattering.

\subsection{X-ray constraints}
\label{sec:xray_constr}
The contact interaction and three-body phase space also permit simple analytical Rayleigh spectra.
We use $x=E_\gamma/\Delta$, with $0<x<1$, for the dimensionless photon energy in the $\chi_1$ rest frame.
At fixed photon energy, the phase-space measure is proportional to $ds$, where $s=k^2$. In the small-splitting limit, the upper boundary is $s_{\max}=4\Delta^2x(1-x)$, giving
\begin{align}
    \frac{dN_{C_S}}{dx} = \frac{dN_{\widetilde C_S}}{dx}
    &= 280\,x^3(1-x)^3
    \,,
    \nonumber\\[2mm]
    \frac{dN_{C_P}}{dx} = \frac{dN_{\widetilde C_P}}{dx} &= 840\,x^3(1-x)^3
    \bigl[1 - 3x(1-x)\bigr]
\,,
\end{align}
where each spectrum integrates to two and has $\langle x\rangle=1/2$.
For $C_P$ and $\widetilde C_P$, the momentum suppression of the pseudoscalar transition bilinear introduces an additional factor $(\Delta^2-k^2)$ in the squared amplitude.
We apply the procedure in \cref{sec:xray_recast} to these spectra, retaining their normalization to two photons per decay.

The operators $\mathcal C_S\in\{C_S,\widetilde C_S\}$ reproduce the contact-limit spectrum of the scalar-mediator model in Ref.~\cite{Krnjaic:2025zjl}. The limit in the right panel of Fig.\,5 of that reference therefore gives
\begin{equation}
    \frac{m_{\chi_0}}{\Gamma_{\mathcal C_S}}>2f_1 L_{\mathcal C_S}(\Delta)
    \,.
    \label{eq:rayleigh_scalar_lifetime_bound}
\end{equation}
At our main benchmark splitting,
\begin{equation}
    L_{\mathcal C_S}(900\,\mathrm{keV}) \simeq 1.04 \times 10^{24}\,\mathrm{GeV}\,\mathrm{s}\,.
    \label{eq:rayleigh_scalar_limit_900}
\end{equation}

\begin{figure}[tbp]
    \centering
    \includegraphics[width=0.45\textwidth]{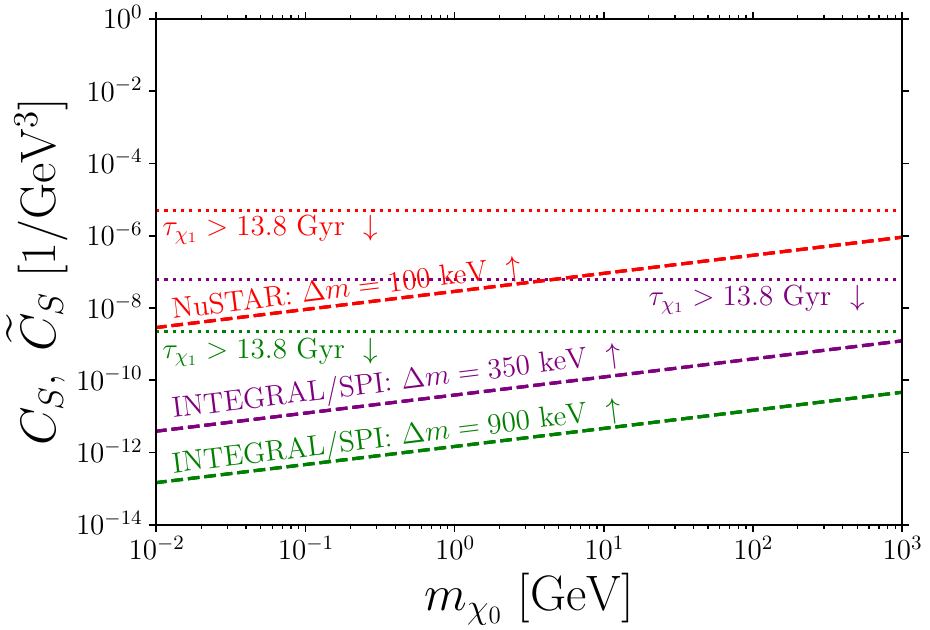}
    
    \vspace{10pt}

    \includegraphics[width=0.45\textwidth]{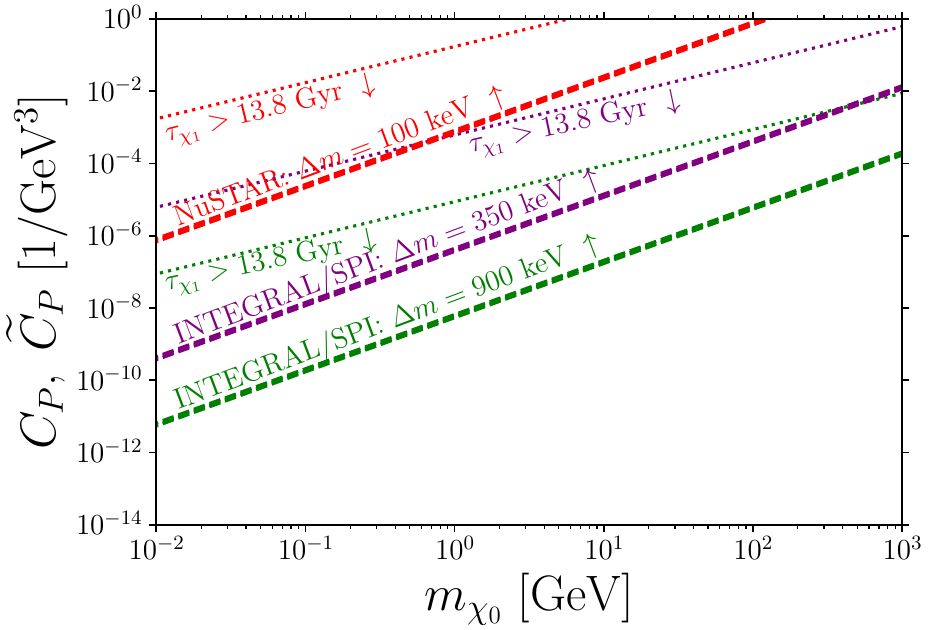}
    \caption{
            Same as \cref{fig:dim6_INTEGRAL_NuSTAR}, but for the Rayleigh operators $C_S,\widetilde C_S$ (top) and $C_P,\widetilde C_P$ (bottom).
            The weaker splitting dependence of the two-photon widths and the additional DM-mass dependence of the pseudoscalar-bilinear operators modify the exclusion bounds and lifetime contours relative to the dimension-6 case. Compare \cref{eq:Gamma_an,eq:G_Rayleigh}.
        }
    \label{fig:Rayleigh_INTEGRAL_NuSTAR}
\end{figure}

The pseudoscalar-bilinear operators $\mathcal C_P\in\{C_P,\widetilde C_P\}$ give a different spectrum with the same mean photon energy, $\langle x\rangle=1/2$, as shown in \cref{fig:spectra}. The unweighted shape overlap and norm ratio are
\begin{equation}
    \cos(P,S)=\sqrt{\frac{153}{160}}\simeq 0.98
    \,,
    \quad
    \left[\frac{\int_0^1dx\,s_P^2(x)} {\int_0^1dx\,s_S^2(x)}\right]^{1/2} \simeq 0.92
    \,.
    \label{eq:rayleigh_flat_recast}
\end{equation}
For INTEGRAL/SPI, the same approximate spectral-recast prescription as for AM gives
\be
    L_{\mathcal C_P}(\Delta) &\simeq 0.92\,L_{\mathcal C_S}(\Delta)
    \,,
\\
    L_{\mathcal C_P}(900\,\mathrm{keV}) &\simeq9.5\times10^{23}\,
    \mathrm{GeV}\,\mathrm{s}
    \,.
    \label{eq:rayleigh_pseudoscalar_lifetime_recast}
\ee
As in the AM case, we assign an estimated $20\%$ recasting uncertainty to the lifetime--mass limit. This accounts approximately for the unavailable energy-dependent covariance and correlations with fitted backgrounds. 

Assuming one operator dominates, the corresponding coefficient bounds are
\begin{align}
    |C_S|,|\widetilde C_S| &\lesssim 1.50\times 10^{-12}\,\mathrm{GeV}^{-3} \left[ \frac{m_{\chi_0}/\mathrm{GeV}} {2f_1 L_{\mathcal C_S,24}(\Delta)} \right]^{1/2}
    \nonumber\\
    &\quad \left(\frac{900\,\mathrm{keV}}{\Delta}\right)^{7/2}
    \,,
    \label{eq:CS_xray_bound}
    \\
    |C_P|,|\widetilde C_P| &\lesssim 5.76\times10^{-9}\,\mathrm{GeV}^{-3} \left[ \frac{(m_{\chi_0}/\mathrm{GeV})^3} {2f_1 L_{\mathcal C_P,24}(\Delta)} \right]^{1/2}
    \nonumber\\
    &\quad \left(\frac{900\,\mathrm{keV}}{\Delta}\right)^{9/2}
\,,
    \label{eq:CP_xray_bound}
\end{align}
where $L_{X,24}= L_X/(10^{24}\,\mathrm{GeV}\,\mathrm{s})$.
We illustrate these limits in \cref{fig:Rayleigh_INTEGRAL_NuSTAR}.

In particular, for $\Delta=900\,\mathrm{keV}$ and $f_1=1/2$, \cref{eq:CS_xray_bound,eq:CP_xray_bound} gives
\be
    |C_S|,|\widetilde C_S| &\lesssim1.47\times10^{-12}\,\mathrm{GeV}^{-3} \sqrt{\frac{m_{\chi_0}}{\mathrm{GeV}}}
    \,,
    \\
    |C_P|,|\widetilde C_P| &\lesssim5.90\times10^{-9}\,\mathrm{GeV}^{-3} \left(\frac{m_{\chi_0}}{\mathrm{GeV}}\right)^{3/2}
\,,
    \label{eq:rayleigh_operator_limits_900}
\ee
which indicates that NuSTAR does not probe the freeze-in region shown in \cref{fig:main_results}.

\bibliography{bibliography}

\end{document}